\documentclass[aps,preprint,pra,10pt,twocolumn,showpacs]{revtex4-1}
\usepackage{amsmath}
\usepackage{graphicx}

\begin{document}
\title{{Multichannel Quantum Defect Theory of Strontium Rydberg Series}}
\author{C L Vaillant}
\email{c.l.j.j.vaillant@durham.ac.uk}
\author{M P A Jones}
\author{R M Potvliege}
\email{r.m.potvliege@durham.ac.uk}
\affiliation{Department of Physics, Joint Quantum Centre (JQC) Durham-Newcastle, Durham University, South Road, Durham DH1 3LE, United Kingdom}
\begin{abstract}
Using the reactance matrix approach, we systematically develop new multichannel quantum defect theory  models for the singlet and triplet S, P, D and F states of strontium based on improved energy level measurements. The new models reveal additional insights into the character of doubly excited perturber states, and the improved energy level measurements for certain series allow fine structure to be resolved for those series' perturbers. Comparison between the predictions of the new models and those of previous empirical and \emph{ab initio} studies reveals good agreement with most series, however some discrepancies are highlighted. Using the multichannel quantum defect theory wave functions derived from our models we calculate other observables such as Land\'e $g_J$-factors and radiative lifetimes. The analysis reveals the impact of perturbers on the Rydberg state properties of divalent atoms, highlighting the importance of including two-electron effects in the calculations of these properties. The work enables future investigations of properties such as Stark maps and long-range interactions of Rydberg states of strontium.
\end{abstract}

\pacs{31.15.-p,31.10.+z,32.70.Cs,32.80.Rm,32.10.Fn, 33.15.Pw, 71.18.+y}

\maketitle
\section{Introduction}
Since its introduction, multichannel quantum defect theory (MQDT) has had much success in the phenomenological description of atomic spectra \cite{Seaton1966,Seaton1983,Aymar1996}. MQDT models have been particularly useful in the analysis of a wide range of Rydberg state properties, including energy levels \cite{Esherick1977}, Land\'e $g$-factors \cite{Wynne1977} and radiative lifetimes \cite{Dai1995a}. 

Recently, cold Rydberg gases have risen to prominence \cite{Saffman2010,Comparat2010}, due to the potential for realising quantum many-body systems \cite{Lukin2001}. In strontium and other two-electron atoms, the extra valence electron adds new possibilities for detection \cite{Millen2010,Lochead2013,McQuillen2013} and trapping \cite{Mukherjee2011,Ovsiannikov2011,Topcu2014}, whereas narrow intercombination lines offer prospects for generating spin squeezed states \cite{Gil2013}.  Recent works \cite{Vaillant2012,Millen2011,Zhi2001} have used a one-electron description to describe the properties of highly excited states of strontium. However, the enhanced spectroscopic resolution now available  using laser-cooled atoms and continuous-wave laser excitation provides a strong motivation for improved modelling that takes two-electron effects into account \cite{Ye2013}.

In previous MQDT studies of strontium, channel interactions were found to cause significant breakdown of $LS$-coupling in the $^{1,3}\mathrm{D}_2$ states \cite{Esherick1977,Wynne1977}. \emph{Ab initio} studies also exist for strontium \cite{Aymar1987,Kompitsas1990,LucKoenig1998}, although most of the models that focussed on bound Rydberg states neglected fine structure effects. The resulting mixing between singlet and triplet states permits the coupling of electric dipole forbidden spin transitions, which may have uses in areas such as research into quantum spin dynamics \cite{Fukuhara2013}. Strontium has attracted less attention in the MQDT community than, for example, barium, where the complex interplay between a multitude of perturbers in many series presented a major challenge for the description of the Rydberg states of barium \cite{Aymar1984,Aymar1996}.

In this article, we systematically develop MQDT models for different Rydberg series in strontium. We begin by reviewing the MQDT formalism and the procedure for extracting energy levels and wave functions from an MQDT model. Following this we develop MQDT models for the $^1\mathrm{S}_0$, $^3\mathrm{S}_1$, $^1\mathrm{P}_1^{\rm   }$, $^3\mathrm{P}_{0,1,2}^{\rm   }$, $^{1,3}\mathrm{D}_2$, $^{3}\mathrm{D}_{1,3}$, $^1\mathrm{F}_3^{\rm   }$ and $^3\mathrm{F}_{2,3,4}^{\rm   }$ series of strontium, including an analysis of the channel interactions in these series. These models are {generally} based on experimental energy level data with smaller uncertainties than those used by previous empirical models. Finally, we provide tests of the resulting MQDT models by calculating Land{\'e} $g_J$-factors for the $^{1,3}\mathrm{D}_2$ series, as well as radiative lifetimes for the $^1\mathrm{S}_0$ and $^{1}\mathrm{D}_2$ series, showing good agreement with experiment over a large range of energies. The accurate calculation of these excited state lifetimes depends crucially on the contribution of doubly excited perturbing states \cite{Bergstrom1986}, which quench the lifetimes of even highly excited Rydberg states they are admixed to. The analysis of perturbers paves the way for the inclusion of two-electron effects in other calculations that require the knowledge of dipole matrix elements, such as the calculation of long-range interactions or the determination of Stark maps.

Atomic units are used throughout the paper, unless otherwise stated. 

\section{MQDT Formulation}
\label{mqdtformulation}
\subsection{The reactance matrix approach}
In the standard formulation of MQDT \cite{Cooke1985,Aymar1996},
 a single {active} electron interacts with the core
 through an effective {core-dependent} interaction potential. Different configurations
of the active electron and the core are referred to as channels.
The theory addresses the case where several different channels coincide in total energy, total angular momentum and parity, so
that the eigenenergies and other properties of the excited states may depend on
their interplay. Of particular interest here are
the properties of Rydberg bound states, which are mostly determined
 by the {Coulomb tail} of the effective {interaction} potential.
 The {radial} wave function
 of the {active}
electron in channel
$i$, {$\phi_i(r)$,}
 satisfies the Coulomb equation at distances $r$ from the nucleus
much larger than, typically, the Bohr radius $a_0$. Namely, one has, 
for $r$ sufficiently large, 
\begin{equation}
  \left[ -\frac{d^2}{dr^2} + \frac{l_i(l_i+1)}{r^2} + \frac{2\mu}{r} \right] \phi_i = 2\mu \epsilon_i \phi_i,
\label{coulombequation}
\end{equation}
where $\mu$ is the reduced mass and $\epsilon_i$ is the energy of the active electron in that channel. As defined by this equation, the energy $\epsilon_i$ is measured relative to the ionization
threshold. However, it is customary in the field to measure energies of atomic 
states relative to the ground state energy and to express them as wave numbers. In the following, we denote by $E$ the total energy of the state and by $I_i$ the ionization limit of channel $i$, with both $E$ and $I_i$ expressed as wave numbers. Thus
\begin{equation}
\epsilon_i = (E - I_i)hc,
\end{equation}
where $h$ is Planck's constant and $c$ the speed of light.

Any solution of equation (\ref{coulombequation}) can be written as a linear superposition of two linearly independent solutions, which we take to be the regular and irregular Coulomb functions, $s(l_i,\epsilon_i,r)$ and $c(l_i,\epsilon_i,r)$, respectively. Both $s(l_i,\epsilon_i,r)$ and $c(l_i,\epsilon_i,r)$ are well known functions which can be calculated numerically using standard algorithms \cite{Seaton2002}. We thus write \cite{Cooke1985}
\begin{equation}
\phi_i(r) = \left[ s(l_i,\epsilon_i,r) \cos \theta_i + c (l_i,\epsilon_i,r) \sin\theta_i \right], \quad 
r \gg a_0.
\label{longrange}
\end{equation}

The angle $\theta_i$ determining the weight of the regular and the irregular Coulomb functions in the superposition can be written in terms of an effective principal quantum number $\nu_i$ such that $\theta_i= \pi \nu_i$. In terms of the principal quantum number, $n$, $\nu_i = n- \delta_i$ where $\delta_i$ is a $n$-dependent quantum defect arising from the effect of the core on the long-range part of the total wave function.

In general, the eigenfunctions of the atomic Hamiltonian are linear superpositions of the different channel wave functions.
Thus, if $\Psi$ is the eigenfunction of a state of interest, one has
\begin{equation}
\Psi= {\cal A}\sum_i A_i \phi_i \chi_i,
\label{totalwavefunction1}
\end{equation}
where the functions $\chi_i$ describe the angular and spin dependence of the active electron
as well as the radial, angular and spin dependence of all the other electrons. The index
$i$ ($i=1,\ldots,N$) runs over all the channels included in the model and the symbol ${\cal A}$
indicates that the sum is antisymmetrized.
The coefficients
$A_i$ determine the admixture of each channel wave function in the eigenstate $\Psi$.
Like the quantum defects $\delta_i$ defined above, these coefficients vary with the
principal quantum number. This dependence on $n$ entirely arises
from interchannel couplings induced by electron-electron interactions
in the vicinity of the nucleus.

At least in principle, one can diagonalize these short-range interactions and pass
to a set
of $N$ linearly independent uncoupled wave functions $\psi_\alpha$ ($\alpha=1,\ldots,N$). These
uncoupled wave functions are related to the channel wave functions $\phi_i\chi_i$ by
a unitary transformation. Without loss of generality, one has,
for $r$ sufficiently large,
\begin{equation}
\begin{split}
\psi_\alpha &= \left( {\cal A}\sum_i U_{i \alpha} s(l_i,\epsilon_i,r) \chi_i \right) \cos(\pi \mu_\alpha)\\
 &- \left({\cal A}\sum_i U_{i \alpha} c(l_i,\epsilon_i,r) \chi_i \right) \sin(\pi \mu_\alpha),
\label{shortrange}
\end{split}
\end{equation}
where the coefficients $U_{i \alpha}$ are real and form the elements of a unitary matrix \cite{Cooke1985}.
(The numbers $\mu_\alpha$ are usually referrred to as eigenquantum defects.)

One can thus write the wave function $\Psi$ in terms of the uncoupled wave
functions, so that 
\begin{equation}
{\cal A}\sum_i A_i \phi_i \chi_i= \sum_\alpha B_\alpha \psi_\alpha.
\label{totalwavefunction}
\end{equation}
(Depending on the state considered and on how the wave functions $\psi_\alpha$ are
defined, some of the constant coefficients $B_\alpha$ appearing in this equation may be zero.)
Taking the limit $r \rightarrow \infty$ in equation (\ref{totalwavefunction}) and
equating the coefficients of $s(l_i,E,r)$ and $c(l_i,E,r)$ separately gives
\begin{subequations}\label{equatingcoefficients}
\begin{align}
A_i e^{-i \pi \nu_i} &= \sum_\alpha U_{i \alpha} B_{\alpha} e^{i \pi \mu_\alpha}\\
B_\alpha e^{i \pi \mu_\alpha} &= \sum_i U_{i \alpha} A_i e^{-i \pi \nu_i}.
\end{align}
\end{subequations}
Requiring the $A_i$ and $B_\alpha$ coefficients to be real leads to a set of equations commonly expressed as
\begin{equation}
\sum_i U_{i \alpha} A_i \sin[ \pi (\nu_i + \mu_\alpha) ]=0, \quad \alpha=1,\ldots,N.
\label{mqdtcondition}
\end{equation}
For a non-trivial solution to exist, one must have
\begin{equation}
\det | U_{i \alpha} \sin[\pi (\nu_i + \mu_\alpha)] | =0.
\label{detcondition}
\end{equation}
For given values of $U_{i\alpha}$ and $\mu_{\alpha}$, this last equation
is a constraint on the effective
principal quantum numbers $\nu_i$.
Introducing the boundary condition that $\Psi \rightarrow 0$ as $r\rightarrow \infty$ leads
to the additional constraints that the total energy must satisfy the equations
\begin{equation}
E=I_i -\frac{\tilde{R}}{\nu_i^2}, \quad i=1,\ldots,N,
\label{rydberg}
\end{equation}
where $\tilde{R}$ is the mass-corrected Rydberg constant for the species considered
and $I_i$ is the ionization limit of channel $i$. The energies $E$ at which the values of $\nu_i$ given by
equation \eqref{detcondition} coincide with those given by equation \eqref{rydberg} are the bound state energies predicted by the model.

The parameters characterizing each series, i.e., the matrix {$[U_{i \alpha}]$} and the eigenquantum defects $\mu_\alpha$, can be calculated \emph{ab initio} \cite{Aymar1987} or by fitting the predicted values of the bound state energies to experimental data using least-squares fitting procedures. One method of obtaining the matrix elements $U_{i \alpha}$ is to write the matrix $[U_{i \alpha}]$ as a product of rotation matrices, where the rotation angles describe the coupling between individual channels. These angles and the eigenquantum defects are then free parameters to be fitted to experimental energy levels. This unitary transformation formulation has been used to develop MQDT models for calcium, strontium and barium \cite{Armstrong1977,Esherick1977,Aymar1984}. However, factorizing the transformation matrix $[U_{i \alpha}]$ into a product of rotation matrices does not lead to a unique representation of $[U_{i \alpha}]$, as the order of the rotations is also required \cite{GiustiSuzor1984}.

Instead of obtaining the matrix elements $U_{i \alpha}$ and the eigenquantum defects $\mu_\alpha$, we found it advantageous to follow \cite{Cooke1985} and rewrite equation \eqref{mqdtcondition} in the form
\begin{equation}
\begin{split}
\cos(\pi \mu_{\alpha}) &\sum_i \left[ U_{i \alpha} \tan(\pi \nu_i) \right. \\
&+  \left. \tan(\pi \mu_\alpha) U_{i \alpha} \right] \cos(\pi \nu_i) A_i =0.
\end{split}
\end{equation}
Given that $[U_{i \alpha}]$ is a unitary matrix, this can also be written
\begin{equation}
\sum_i \left[ K_{i \alpha} + \delta_{i \alpha} \tan(\pi \nu_i) \right] a_i = 0,
\label{rmatrixmqdtcondition}
\end{equation}
where $a_i =  A_i \cos(\pi \nu_i)$ and the coefficients $K_{i \alpha}$ are the elements of
a real symmetric matrix $[K_{i\alpha}]$ defined by the equation 
\begin{equation}
[K_{i \alpha}] = [U_{i\alpha}]^\dagger [\delta_{\alpha i} \tan(\pi \mu_\alpha)] [U_{i \alpha}].
\label{kmatrix}
\end{equation}
The condition equivalent to equation \eqref{detcondition} is then given by
\begin{equation}
\det | K_{i \alpha} + \delta_{i \alpha}\tan(\pi \nu_i) | = 0.
\label{collisiondetcondition}
\end{equation}

In this formulation of the theory, the coupling between the different channels is entirely described by the real symmetric matrix $[K_{i \alpha}]$, which is known as the reactance matrix. This formulation is well suited to numerical computations using standard optimization routines, both for calculating bound state energies and for calculating the matrix elements $K_{i \alpha}$ by fitting these energies to experimental data.

It is worth noting that effects such as core polarization may lead to a dependence of the reactance matrix on the energy of the states \cite{Seaton1983}. This variation is often negligible over the range of energy considered. Occasionally, however, this is not the case, and it is found that the bound state energies cannot be satistfactorily reproduced over a whole Rydberg series with constant values of the matrix elements $K_{i \alpha}$. In this work, we account for this energy dependence by allowing the diagonal elements of the reactance matrix to vary linearly with the energy, $E$. We write \cite{Esherick1977}
\begin{equation}
K_{i i} (E) = K_{i i}^{(0)} + K_{i i}^{(1)}  \frac{(I_s - E)}{I_s},
\label{energydependence}
\end{equation}
where $I_s$ is the first ionization threshold ($I_s = 45 \, 932.1982 \; \mathrm{cm}^{-1}$ in strontium \cite{Beigang1982a}) and $K_{i i}^{(0)}$ is the value of the matrix element $K_{ii}$ at $E=I_s$. The coefficients $K_{i i}^{(0)}$ and $K_{i i}^{(1)}$ are fitting parameters.

\subsection{Bound state energies}
\label{boundstatesection}
As already mentioned, the bound state energies predicted by the MQDT model are those values of $E$
for which equations  \eqref{rydberg} and \eqref{collisiondetcondition} are satisfied simultaneously. 
We proceed as follows to find these energies. To start, we select two of the $N$ channels,
for example channel $j$ and channel $k$, and assign them specific roles in the calculation. (We take for
channels $j$ and $k$ the ones which are expected to dominate the series
analysed; however, this choice is largely arbitrary and we have verified that different
choices lead to the same final results.) We set the effective principal quantum
number $\nu_j$ to a value guessed from experimental data and use
equation  \eqref{rydberg} to derive the corresponding values of all the other quantum numbers
$\nu_j$, apart from $\nu_k$.
Namely, we introduce a function $F_i(\nu_j)$ defined by the equation
\begin{equation}
F_i(\nu_j)= \left[ \frac{I_i - I_j}{\tilde{R}} + \frac{1}{\nu_j^{2}} \right]^{-{1}/{2}}
\end{equation}
and set
\begin{equation}
\nu_i = F_i(\nu_j), \qquad i\not=k.
\label{rydbergconvert}
\end{equation}
We use equation \eqref{collisiondetcondition}
to determine the value of $\nu_k$ as an implicit function of the value
of $\nu_j$ for given values of the matrix elements $K_{i\alpha}$.
Thus, we define the function $G_k([K_{i\alpha}];\nu_j)$ such that
equation \eqref{collisiondetcondition} is fulfilled when $\nu_k=G_k([K_{i\alpha}];\nu_j)$ and 
$\nu_i=F_i(\nu_j)$, $i\not=k$ (see appendix \ref{appendixA} for a two-channel example).
We then seek the values of
$\nu_j$ for which the values of $\nu_k$ so obtained are consistent with
equation \eqref{rydberg}; that is, we find the zeros of the function $\Xi(\nu_j)$, where
\begin{equation}
\Xi(\nu_j) = G_k([K_{i\alpha}];\nu_j)-F_k(\nu_j).
\label{difference}
\end{equation}
The resulting set of self-consistent effective principal quantum numbers
gives, through equation \eqref{rydberg}, the theoretical bound state spectrum
defined by the reactance matrix $[K_{i\alpha}]$. We calculate the elements of this matrix 
by $\chi^2$-fitting the theoretical energies to spectroscopic data for the
series considered.

As the poles of the functions $\tan(\pi\nu_i)$ complicate the automated location of the zeros of $\Xi(\nu_j)$, we look instead for the minima
of the function $\Xi(\nu_j)^2$, which is a simpler computational problem.
 We use a standard Brent's method minimization algorithm to
this effect \cite{Scipy2001}.
The algorithm is started with an initial guess based on the experimental values of $\nu_{j}$,
with the upper and lower bounds being suitable fixed values above and below this experimental
value. This determines the energy levels to roughly six significant figures,
depending on the particular energy level being determined.
To find the best MQDT model, the resulting bound state energies are then used to calculate
a value of $\chi^2$, where
\begin{equation}
\chi^2= \sum_n [(E^{(n)}_{\mathrm{exp}} - E^{(n)}_{\mathrm{MQDT}})/\alpha^{(n)}_{\mathrm{exp}}]^2.
\label{chisquared}
\end{equation}
Here, $E^{(n)}_{\mathrm{MQDT}}$, $E^{(n)}_{\mathrm{exp}}$ and $\alpha^{(n)}_{\mathrm{exp}}$ are,
respectively, the MQDT prediction for the energy of the state with principal quantum number
$n$, the measured value of this energy, and the experimental error on this measured value.
We use a Nelder-Mead simplex algorithm \cite{Scipy2001} to find the values of the matrix
elements $K_{i \alpha}$ minimizing the value of $\chi^2$.
The quality of the fit is characterized by the 
reduced $\chi^2$ value, $\chi^2_\nu= \chi^2/\nu$, where $\nu$ is the number
of degrees of freedom (not to be confused with an effective principal quantum number $\nu_i$).
Assuming that the scatter in the data is random and normally distributed,
the model can be considered to match the data well when 
$\chi^2_\nu \approx 1$ \cite{Hughes2010}.

\subsection{Wave functions and channel fractions}
\label{fractionssection}
Besides the bound state energies, the mixing coefficients {$A_i$ and $B_{\alpha}$} can
also be determined once the matrix elements $K_{i\alpha}$ have been obtained.
Through this, any required observable can be calculated, such as dipole matrix elements.

The simplest way to obtain the full {wave function} $\Psi$ is by working with the channel wave functions $\phi_i(r)\chi_i$. To calculate the coefficients $A_i$, we solve the matrix equation \eqref{rmatrixmqdtcondition}. This yields a set of $N$ simultaneous equations which determine the $A_i$ coefficients to within an arbitrary overall factor (recall that $a_i=A_i\cos(\pi\nu_i)$). Taking
\begin{equation}
\int^\infty_0 |\phi_i(r)|^2 dr =1,
\label{channelnormalization}
\end{equation}
we adopt the normalization prescription of  \cite{Fano1970}, namely
\begin{equation}
\sum_i \bar{A}_i^2 = 1
\label{eq:normal}
\end{equation}
with
\begin{equation}
\bar{A}_i = \nu_i^{3/2} A_i.
\end{equation}
Using this normalization in combination with equation
\eqref{rmatrixmqdtcondition} leads to a fully determined system.
The coefficients $A_i$ can thus be determined once the values of the effective
principal numbers $\nu_i$ and of the matrix elements $K_{i \alpha}$
have been obtained (see appendix \ref{appendixA} for a two-channel example).

We follow this approach to obtain the channel fractions $\bar{A}_i^2$ needed to calculate Land{\'e} $g_J$-factors and radiative lifetimes
of highly excited states. The dipole matrix elements involved in the calculation of the
radiative lifetimes are obtained by approximating the radial channel wave functions
$\phi_i(r)$ by their large-$r$ form, Eq. \eqref{longrange}, with the Coulomb functions
$s(l_i,E,r)$ and $c(l_i,E,r)$ calculated using procedures
detailed in  \cite{Seaton2002}. As is explained in section \ref{lifetimesection}, we
do not use the (unknown) radial wave functions of the core electrons for obtaining the
dipole matrix elements of interest.

\subsection{Spin-orbit effects}
A $jj$-coupling scheme is the natural framework when
different channels converge to ionization limits split by the spin-orbit interaction.
In many cases, however, the splitting between the fine structure components
is small enough that a simple average of the ionization limits can be taken,
which makes it possible to develop the whole calculation in $LS$-coupling. (In this work,
and following \cite{Esherick1977}, the average is an unweighted arithmetic average
of the fine structure components.)

This approach may break down when two channels of the same configuration
but different fine structure components are coupled to each other,
as the $L$ and $S$ quantum numbers labelling the different channels across the series may then 
become ambiguous.
This situation lends itself to a $jj$-coupling scheme,
with the resulting wave functions being $jj$-coupled rather than $LS$-coupled.
In this case assigning the different
components of the wave functions to particular $LS$ quantum numbers can be done
by recoupling the $jj$-coupled wave functions using standard Racah algebra \cite{Brink1993}:
\begin{equation}
\begin{split}
{\bar A}_i^{(LS)} &= \sum_{i'} \delta_{l^{}_{1i'} l^{}_{1i}}\delta_{l^{}_{2i'} l^{}_{2i}}\delta_{s^{}_{1i'} s^{}_{1i}}\delta_{s^{}_{2i'} s^{}_{2i}}\\
&\times\sqrt{(2j^{}_{1i'}+1)(2j^{}_{2i'}+1)(2L^{}_i+1)(2S^{}_i+1)} \\
&\times\begin{Bmatrix}
l^{}_{1i'} & s^{}_{1i'} & j^{}_{1i'}\\
l^{}_{2i'} & s^{}_{2i'} & j^{}_{2i'}\\
L^{}_i & S^{}_i & J^{}
\end{Bmatrix} {\bar A}_{i'}^{(jj)},
\end{split}
\label{recoupling}
\end{equation}
where the subscripts 1 and 2 refer to each of the two valence electrons and 
where $J$ is the total angular momentum quantum number (which is the same
for all the $jj$-coupled wave functions included in the sum, as channels with
different values of $J$ do not interact with each other).
The quantities ${\bar A}_i^{(LS)2}$ and ${\bar A}_{i'}^{(jj)2}$, $i,i'=1,\ldots,N$,
 are the channel fractions for, respectively, the $LS$-coupled and $jj$-coupled channels. 

\section{Bound state energies of strontium}
\label{SectionIII}
In this section we present the details of our MQDT models of the S, P, D and F series
of strontium and compare our results to 
previous work. 
In empirical MQDT models, such as those obtained here, 
the MQDT parameters are fitted to reproduce experimental energy levels. By contast, 
in \emph{ab initio} models, such as those derived from $R$-matrix calculations, these parameters are obtained without direct input
from experiment \cite{Aymar1996}. 
The \emph{ab initio} approach allows for a more detailed understanding of the interactions 
between the valence electrons. The resulting models, however, are rarely capable of reproducing 
experimental data to within the
measurement uncertainties. Empirical models are thus useful for refining
\emph{ab initio} results and obtaining accurate
theoretical energy levels. 
However, they have a limited range of applicability; 
we find, for instance, that in strontium bound state energy levels
below about $38\, 000 \; \mathrm{cm}^{-1}$ are 
difficult to reproduce due to {non-linear energy dependences that cannot be taken into account 
through equation \eqref{energydependence}.}

\begin{figure*}[tb]
\centering
\includegraphics{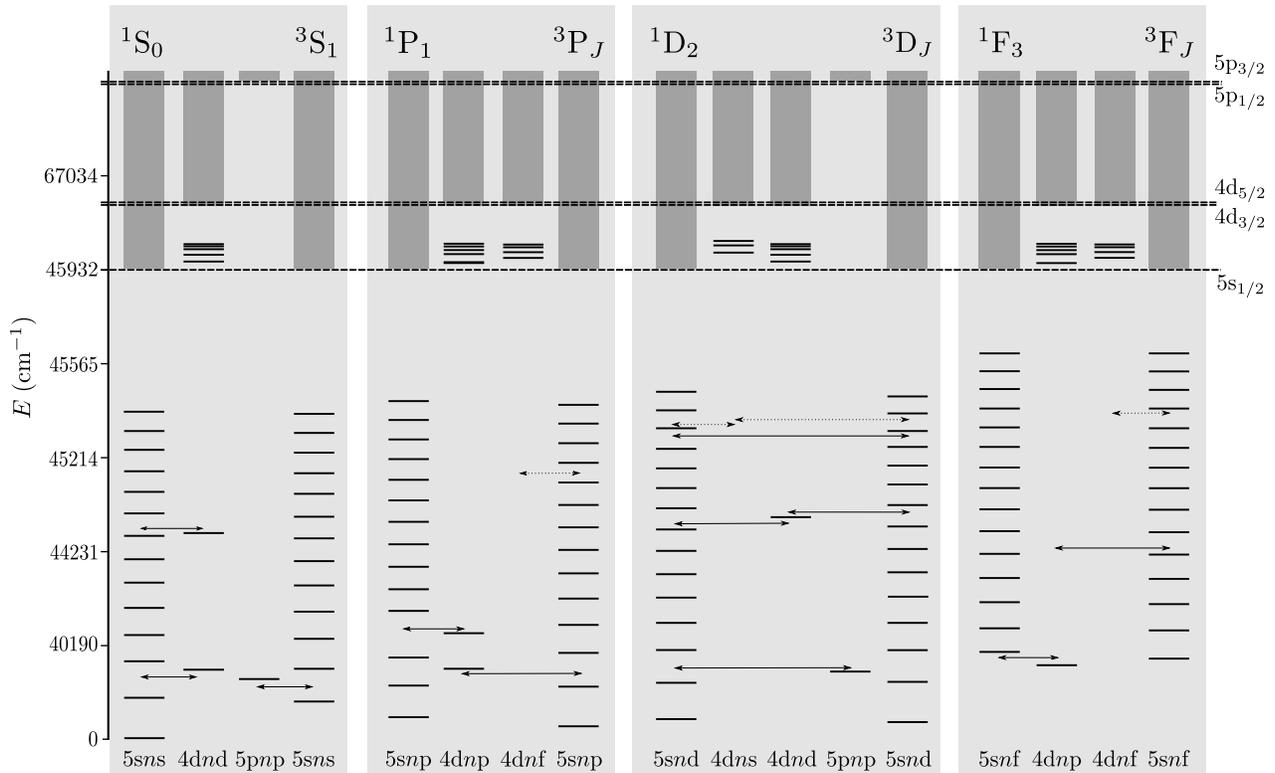}
\caption{Schematic of the S, P, D and F series of strontium, indicating
the interaction between channels.  The dashed lines mark the ionization limits, as
identified on the right-hand side of the figure by the state of the residual ion. 
The solid lines represent the positions of the bound states in each series for the different
configurations considered, in cm$^{-1}$, neglecting
fine structure. The grey boxes represent the
respective continua. The vertical scale is non-linear. 
The solid arrows indicate a perturbation between channels where an existing bound state can be identified as the relevant perturber. The dashed arrows indicate perturbations where the perturber is not a distinct state but is instead mixed over a range of bound states. The terms of the doubly excited
states are not given in the diagram and may differ
from those of the series they perturb.} 
\label{seriesoverview}
\end{figure*}

A schematic of the series considered in this work, showing how the various channels interact,
is given in figure~\ref{seriesoverview}.
As indicated by this figure, the perturbers are generally doubly
excited states formed by combinations of low lying orbitals, namely the 5s, 6s, 5p, 4d and 4f orbitals. (The lowest energy states of 
the $\mathrm{Sr}^+$ ion are, in order of increasing energy, the $5\mathrm{s}_{1/2}$, 
$4\mathrm{d}_{3/2}$, $4\mathrm{d}_{5/2}$, $5\mathrm{p}_{1/2}$, $5\mathrm{p}_{3/2}$ and 
$6\mathrm{s}_{1/2}$ states \cite{Sansonetti2012}; the lowest states of f symmetry, namely the $4\mathrm{f}_{7/2}$ and 
$4\mathrm{f}_{5/2}$ states, lie above the  $6\mathrm{p}_{3/2}$ state.)
The way the corresponding channels interact with each other varies from series to series, as channels significantly mix only if they have bound states of same parity and total angular momentum lying sufficiently close in energy. Doubly excited states with the same $L$ and $S$ quantum numbers as the 5s$nl$ series they interact with tend to produce a greater perturbation than those with different quantum numbers --- for example, the 4d$^2$ $^1$S$_0$ state perturbs far more the  5s$n$s $^1$S$_0$ series than the  4d$^2$ $^3$P$_0$-even state does. As is well known, the existence of doubly excited states where the two valence electrons are in the same $nl$ orbital is conditional to $L+S$ being even, due to the Pauli exclusion principle \cite{Cowan1981}. Thus, while the 5s$n$s $^1$S$_0$ series is perturbed by the 4d$^2$ $^1$S$_0$ state, the  5s$n$s $^3$S$_1$ series is not perturbed by a 4d$^2$ $^3$S$_1$ state, which is not realisable (figure 1).

The parameters of the MQDT models obtained in this work are tabulated in appendix \ref{models}. In the rest of this section we give details of the channel fraction results obtained using the new MQDT models, and briefly review the results of previous studies, from both \emph{ab initio} and empirical investigations.

\subsection{$5\mathrm{s}n\mathrm{s}\; ^{1}\mathrm{S}_0$ states}
\label{sectionsingletS}
Previous studies of the 5s$n$s $^1\mathrm{S}_0$ series of strontium showed that an empirical two-channel MQDT model satisfactorily reproduced the data available at the time \cite{Esherick1977,Dai1995a,Wynne1979}. They also showed that this series is perturbed by a bound state lying between the 5s6s and 5s7s states, at $37 \, 160.234$ {cm}$^{-1}$ \cite{Sansonetti2010,Moore1952}. These previous studies, however, disagreed as to which two channels should be chosen for the fitting of the $^1\mathrm{S}_0$ states: choosing either a $5\mathrm{p}n\mathrm{p}$ or a $4\mathrm{d}n\mathrm{d} $ perturbing channel results in an equally satisfactory fit for the $5\mathrm{s}n\mathrm{s}$ series. The ambiguity of the character of the perturbing channel was later explained by multi-configuration Hartree-Fock (MCHF) calculations \cite{FroeseFischer1981,Aspect1984}, which found that the admixture of the $4\mathrm{d}n\mathrm{d}$ configuration in the 5s6s and 5s7s states and in the perturber is similar in magnitude, if not almost equal, to the admixture of the  $5\mathrm{p}n\mathrm{p}$ configuration. The importance of the 4d$^2$ configuration in these low lying states has been confirmed by an analysis of their isotope shift \cite{Aspect1984}. An eigenchannel $R$-matrix calculation of this series, while not conclusive as to the character of the perturber, pointed towards a predominant 4d$^2$ character \cite{Aymar1987}. This perturber, however, is listed as a 5p$^2$ $^1\mathrm{S}_0$ state both in \cite{Sansonetti2010} and \cite{Moore1952}, although in \cite{Sansonetti2010} the same term and configuration are also assigned to an autoionizing state at $54\, 451.0$ {cm}$^{-1}$ (the latter assignment is consistent with \emph{ab initio} calculations which showed that this autoionizing state has indeed a predominantly $5\mathrm{p}^2 \; ^1\mathrm{S}_0$ character \cite{Kompitsas1991,LucKoenig1998}). As a bound state generally identified with the $4\mathrm{d}^2 \; ^3\mathrm{P}_0$-even state lies at $E= 44\, 525.838$ {cm}$^{-1}$ \cite{Esherick1977,Sansonetti2010}, the perturber at 37\, 160.234 cm$^{-1}$ is probably best referred to as the $4\mathrm{d}^2 \; ^1\mathrm{S}_0$ state. (The MQDT analysis does not give information on the term symmetry of this perturber, apart that it is an even $J=0$ state.)
 
The $^1\mathrm{S}_0$ series has now been measured precisely over a large range of principal quantum numbers, typically with error bars of the order of $0.001 \; \mathrm{cm}^{-1}$ \cite{Beigang1982a,Sansonetti2010}. The considerably higher precision of these results compared to those the MQDT calculations of \cite{Esherick1977,Dai1995a,Wynne1979} were based on makes it useful to revisit this series.  To this end, we considered two different three-channel MQDT models. In the first one, and on the basis of the \emph{ab initio} results, we included the 5s$n$s, 5p$n$p and 4d$n$d channels, with the 5p$n$p and 4d$n$d channels being taken to converge to the average of the two fine-structure components of the respective ionization limits. This model is unsatisfactory, however, in that it predicts a 5p$n$p character for the $4\mathrm{d}^2 \; ^3\mathrm{P}_0$-even state at $E= 44\, 525.838$ {cm}$^{-1}$ state and in that it yields spontaneous lifetimes in disagreement with experiment. Our second model, which is more successful, includes the fine-structure resolved $jj$-coupled $4\mathrm{d}_{3/2}n\mathrm{d}_{3/2}$ and $4\mathrm{d}_{5/2}n\mathrm{d}_{5/2}$ $J=0$ channels besides the $5\mathrm{s}_{1/2}n\mathrm{s}_{1/2}$ channel.  

We include all the $5\mathrm{s}n{s} \; ^1\mathrm{S}_0$ states with $n$ between 7 and 30 in the MQDT fit, as well as the $4\mathrm{d}^2 \; ^3\mathrm{P}_0$ state at $44\, 525.838$ {cm}$^{-1}$. We use the spectroscopic data of \cite{Beigang1982a,Sansonetti2010}. We did not include the $5\mathrm{s}n{s} \; ^1\mathrm{S}_0$ states with $n > 30$ since their experimental quantum defects show deviations from the expected smooth variation, as discussed in \cite{Beigang1982a,Vaillant2012}. As described in the previous section, we generate the MQDT model using a $\chi^2$ fitting procedure to find the reactance matrix that best reproduces the experimental energy levels. The MQDT parameters for the $jj$-coupled model are presented in appendix \ref{models}. With these parameters, $\chi_\nu^2=0.38$ and the rms deviation is $6 \times 10^{-4} \; \mathrm{cm}^{-1}$ for the singlet series, excluding the low lying bound state at $E= 37 \, 160.234 \; \mathrm{cm}^{-1}$ from the fit.

\begin{figure}[tb]
\centering
\includegraphics{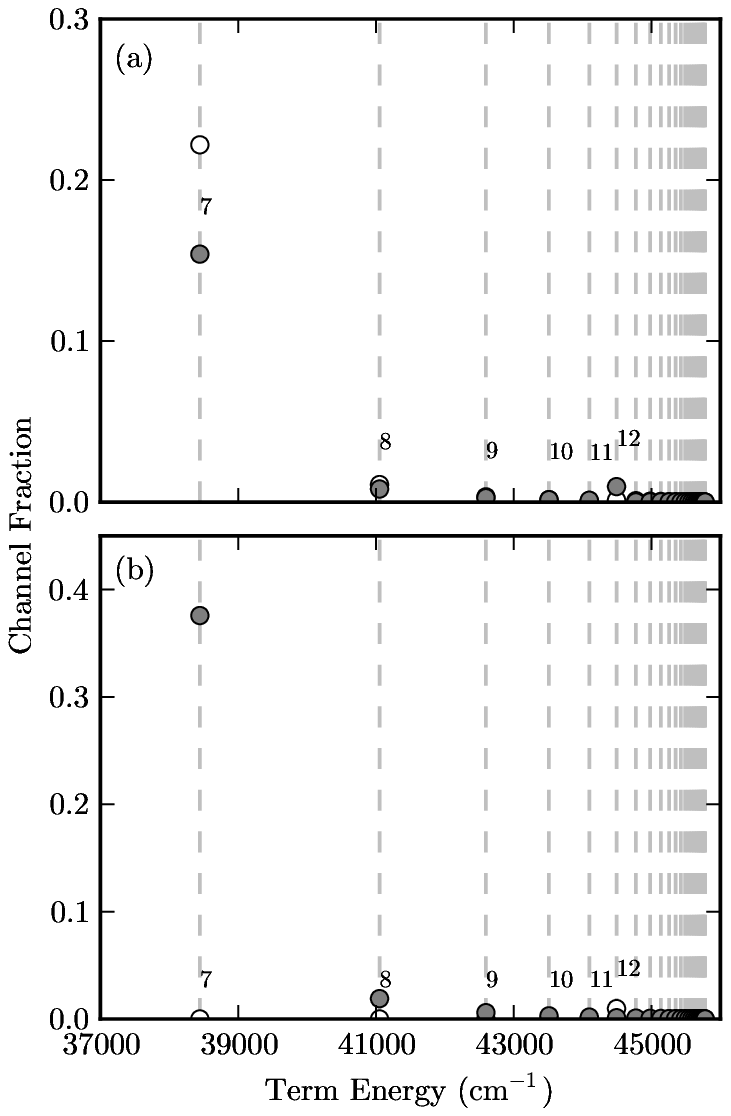}
\caption{Channel fractions of the perturbing channels in the 5s$n$s $^1\mathrm{S}_0$ series of strontium ($7\le n \le 30$). In panel (a), the grey and the white circles refer, respectively, to the $4\mathrm{d}n\mathrm{d}$ $^3\mathrm{P}_0$-even and $4\mathrm{d}n\mathrm{d}$ $^1\mathrm{S}_0$ $LS$-coupled channels, 
while in panel (b) they refer, respectively, to the $4\mathrm{d}_{3/2}n\mathrm{d}_{3/2}$ and $4\mathrm{d}_{5/2}n\mathrm{d}_{5/2}$ $J=0$ $jj$-coupled channels. The vertical dashed lines indicate the experimental positions of the 5s$n$s bound state energies.}
\label{singletsstatefraction}
\end{figure}

The channel fractions of the $4\mathrm{d}_{3/2}n\mathrm{d}_{3/2}$ and $4\mathrm{d}_{5/2}n\mathrm{d}_{5/2}$ $J=0$ channels can be converted into $LS$-coupled
$^1$S$_0$ and $^3$P$_0$ channel fractions using equation \eqref{recoupling}.
These $LS$-coupled channel fractions for the singlet S states are shown in panel (a) of figure \ref{singletsstatefraction}, and the $jj$-coupled channel fractions in panel (b).
The series can be seen to be  strongly perturbed below $n=8$. The  channel fractions in the $jj$-coupled model indicate 
that this perturbation arises
from an interaction between the 5s$n$s and $4\mathrm{d}_{3/2}n\mathrm{d}_{3/2}$ 
channels. In terms of $LS$-coupled channels, both   
the  $^1\mathrm{S}_0$ and $^3\mathrm{P}_0$ symmetries contribute
almost equally. The importance of the 4d$n$d configuration in the low-$n$ end
of the series is borne out by the results of the (5s$n$, 5p$n$p, 4d$n$d) MQDT
model, not shown in figure \ref{singletsstatefraction}, within which the 4d$n$d
channel fraction increases rapidly when $E$ decreases below 
42\, 000 cm$^{-1}$ whereas the
5p$n$p fraction remains close to zero. 

Unfortunately, the low lying perturber could not be reproduced with either one of the two MQDT models, which prevented an unambiguous assignment of this state (including the perturber in the fit made its theoretical energy deviate from its
experimental position by about $3 \; \mathrm{cm}^{-1}$). However, 
we found no evidence for a significant admixture
of the 5p$n$p configuration in the low lying 5s$n$s  $^1$S$_0$ states.
We note that this result is at variance with the predictions of the MCHF calculations
\cite{FroeseFischer1981,Aspect1984,Vaeck1988}.

A smaller perturbation is seen in the vicinity of the $4\mathrm{d}^2$ $^3$P$_0$ state ($E=44\, 525.838$ {cm}$^{-1}$).
Here it is the 4d$_{5/2}n$d$_{5/2}$ channel which gives the largest
contribution. In terms of $LS$-coupled channels, the
$^3$P$_0$ channel fraction is more enhanced than the  $^1$S$_0$ one, whose increase is invisible on the scale of the figure.
The perturbation is small, though, and is only uncovered when fitting the model to precise experimental energies \cite{Beigang1982a}.

\begin{figure}[tb]
\centering
\includegraphics{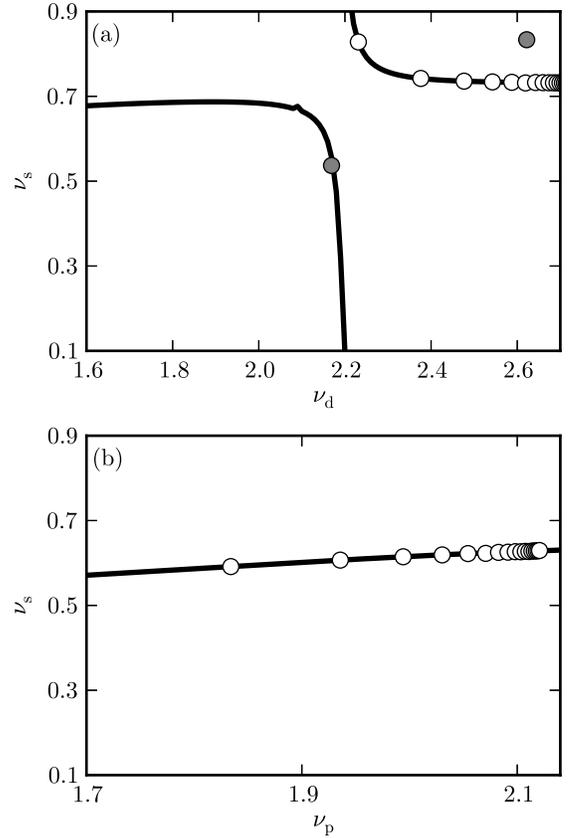}
\caption{Lu-Fano plot for the 5s$n$s $^1\mathrm{S}_0$ series (panel (a)) and for the 5s$n$s $^3\mathrm{S}_1$ series (panel (b)). The open circles represent the experimental positions of the $5\mathrm{s}n\mathrm{s}$ bound states, the grey circles those of the  perturbers. $\nu_\mathrm{s}$ is calculated relative to $I_\mathrm{s}= 45 \, 932.1982 \; \mathrm{cm}^{-1}$, $\nu_\mathrm{p}$ relative to the average of the fine structure components of the $5\mathrm{p}$ ionization limits, $I_\mathrm{p}= 70 \, 048.11 \; \mathrm{cm}^{-1}$, and $\nu_\mathrm{d}$ relative to the 4d$_{3/2}$ ionization limit,  $I_{{\rm d}_{3/2}}= 60 \, 488.09 \; \mathrm{cm}^{-1}$.}
\label{sstatelufano}
\end{figure}

The panel (a) of figure \ref{sstatelufano} shows a Lu-Fano plot for the singlet series. The obvious resonance feature visible at $\nu_{\rm d} \approx 2.2$ arises from the perturbation of the singlet states by the low-lying perturber. For clarity, a narrow resonance feature  at $\nu_{\rm d} \approx 2.62$ arising from the $4\mathrm{d}^2$ $^3$P$_0$ perturber has not been represented in the figure.

\subsection{{$5\mathrm{s}n\mathrm{s}\; ^{3}\mathrm{S}_1$} states}
It seems that no MQDT analysis of the {$5\mathrm{s}n\mathrm{s}\; ^{3}\mathrm{S}_1$} series
has been published so far. However, theoretical one-channel quantum defects have been calculated
and compared to experiment
\cite{Aymar1987,Beigang1982b}. This previous work found no evidence that
this series is perturbed. An MQDT analysis of the measured energies is nonetheless
possible. We have considered two empirical two-channel models in this work, namely
one which includes the 5p$n$p channel besides the 5s$n$s channel, and
one which includes the 4d$n$d channel instead. In both cases, the perturbing channel is taken
to converge to the fine-structure average of the respective ionization limits. The configurations of 
the perturbing channels were chosen in view of the presence of a 5p$^2$  $^3\mathrm{P}_1$
state at 35\, 400.105 cm$^{-1}$ and a 4d$^2$  $^3\mathrm{P}_1$
state at 44\, 595.920 cm$^{-1}$ \cite{Sansonetti2010}, which could, in principle, perturb
the {$5\mathrm{s}n\mathrm{s}\; ^{3}\mathrm{S}_1$} series.

Unfortunately, the data are somewhat less favourable for the triplet S states than for the singlet ones, 
the experimental error bars ranging from $0.01$ to $0.35 \; \mathrm{cm}^{-1}$ \cite{Sansonetti2010,Kunze1993,Beigang1982b}. For both of our models, the fit includes all the 5s$n$s states with $7 \le n \le 23$.  \cite{Beigang1982b} gives triplet energy levels to higher values of $n$; however the corresponding quantum defects depart from the expected smooth variation above $n \approx 20$, and we have therefore excluded them from the calculation. The resulting (5s$n$s, 5p$n$p) MQDT model fits the data with $\chi_\nu^2=0.14$; however, we could not find a good fit with the (5s$n$s, 4d$n$d) model.

\begin{figure}[tb]
\centering
\includegraphics{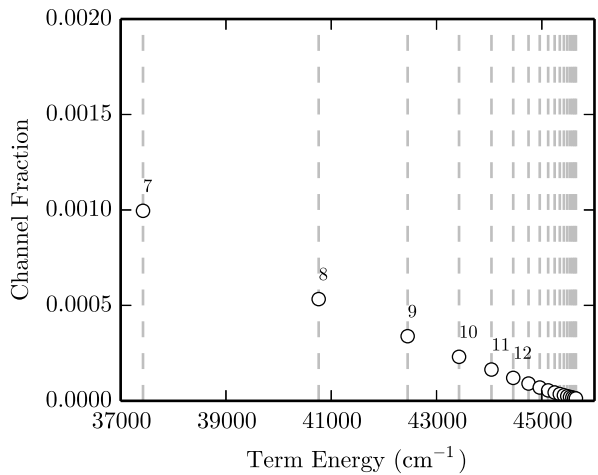}
\caption{Channel fraction of the 5p$n$p channel in the 5s$n$s $^3\mathrm{S}_1$ series of strontium. The vertical dashed lines indicate the experimental positions of the 5s$n$s bound state energies.}
\label{tripletsstatefraction}
\end{figure}

A Lu-Fano plot for the (5s$n$s, 5p$n$p) model is shown in panel (b) of figure \ref{sstatelufano}. An unperturbed series would be flat over the whole range of energies below the first ionization limit. The triplet states are indeed flat, apart from a slope originating from the energy dependence of the diagonal elements of the reactance matrix. The absence of any resonance structure in this plot is in agreement with the conclusion of \cite{Aymar1987,Beigang1982b} that this series is unperturbed. The results shown in figure \ref{tripletsstatefraction}, however, indicate a small but not zero admixture of the 5p$n$p channel in the low-lying 5p$n$s $^{3}\mathrm{S}_1$ states, perhaps originating from a weak coupling between these states and the 5p$^2$  $^3\mathrm{P}_1$ state. There is no sign of a perturbation at the position of the 4d$^2$  $^3\mathrm{P}_1$ state. However, less can be said about how the channels interact in triplet states than in the singlet states because of the lower quality of the data.

\subsection{$5\mathrm{s}n\mathrm{p} \; ^{1}\mathrm{P}_1^{\rm  o}$ states}
The singlet P$^{\rm   }$ states of strontium could be considered a prototypical example of the success of MQDT, as 
this series is  described to a high degree of accuracy by a simple two-channel model \cite{Esherick1977,Lu1977,Aymar1987}. However, the spectroscopic data of \cite{Esherick1977}, on which these previous MQDT analyses were based, have since been superseded by improved measurements \cite{Sansonetti2010}, which has motivated us to reanalyse this series. Like previous investigators, we use a two-channel model including the 5s$n$p and 4d$n$p channels, and include the 4d5p $^1$P$_1$ state in the fit besides the 5s$n$p $^1$P$_1$ states. We find that the higher precision of the new data only helps to reduce the uncertainties on the parameters of the model and this new analysis does not reveal new information on the interaction between the channels. The channel fractions and a Lu-Fano plot for this series can be found, respectively, in the panels (a) of figures \ref{pstatefraction} and \ref{pstatelufano}. The perturbation of the series by the 4d5p $^1$P$_1$ bound state (at 41\,172.054 cm$^{-1}$ \cite{Sansonetti2010}) manifests as a broad resonance structure in both figures.

\subsection{$5\mathrm{s}n\mathrm{p} \; ^{3}\mathrm{P}_0^{\rm  o}$, $^{3}\mathrm{P}_1^{\rm  o}$ and $^{3}\mathrm{P}_2^{\rm  o}$ states}
\label{pstates}
In contrast to their singlet counterparts, the triplet P$^{\rm   }$ series are relatively poorly known.
Spectroscopic data are scarce for these states, due to the difficulty of experimentally 
accessing them. While precise energy levels are available for $n=5$, 6 and 7 
\cite{Sansonetti2010}, the data for $n>7$ we are
aware of \cite{Armstrong1979} are more imprecise. For these highly excited states, 
the experimental single-channel quantum defect of the $J=0$, 1 and 2 series
show significant departures from the expected 
smooth variation with principal quantum number, the $n=13$ energy level being particularly
aberrant for all three series. The largest 
number of energy levels have been measured for the $^3\mathrm{P}_2^{\rm   }$ series; however, the experimental quantum defects oscillate as $n$ increases, 
perhaps due to systematic errors unaccounted for. The $^3\mathrm{P}_1^{\rm   }$ and 
$^3\mathrm{P}_0^{\rm   }$ states also show large scatter in the quantum defects, 
although the data are so few that any trend is difficult to discern. 
 
Reference \cite{Armstrong1979} gives an empirical two-channel MQDT treatment of the triplet series, 
including the $4\mathrm{d}n\mathrm{p}$ channel besides the $5\mathrm{s}n\mathrm{p}$ channel. However,
this model deviates significantly from the  experimental energies, the calculated energies being many standard deviations away from the data.
Nonetheless, an \emph{ab initio} analysis \cite{Aymar1987} showed that these two
channels interact in the triplet series, as they do in the singlet series. (The fine structure of
these triplet states was
not resolved in this \emph{ab initio} study.)
We have reanalysed the data of  \cite{Sansonetti2010} and \cite{Armstrong1979}, including all the triplet 5s$n$p states with $n > 5$ in the fit 
as well as the 4d5p $^3$P$_0$,  $^3$P$_1$ and $^3$P$_2$ 
states. (The $5\mathrm{s}5\mathrm{p}$ state is too low in energy for an MQDT analysis.) We found
that a two-channel MQDT model including only the 5s$n$p channel and a channel converging to the
fine-structure averaged $4\mathrm{d}$ ionization limit is
insufficient to describe the data to within their error. On the other hand,
adding a second channel converging to the 
$4\mathrm{d}$ limit to this model makes it possible to
describe
all three components of the triplet states to within the experimental uncertainty 
if the $n=13$ level is excluded from the calculation. Including the $n=13$ level prevents a good fit, 
$\chi^2_\nu$ being as high as 7.9, 3.2 and 5.6 for the $^3\mathrm{P}_0^{\rm   }$, $^3\mathrm{P}_1^{\rm   }$ and 
$^3\mathrm{P}_2^{\rm   }$ symmetries respectively for $6 \le n \le 15$.

Multiconfiguration Hartree-Fock calculations have shown that in the singlet series there 
is a small admixture of $4\mathrm{d}4\mathrm{f} \; ^1\mathrm{P}_1^{\rm   }$ \cite{Vaeck1988}. 
Whereas this admixture is too small to be resolved in an empirical MQDT analysis of this series,
the most likely configurations of the two 4d$nl$ channels included in our model for the
triplet states are 4d$n$p and 4d$n$f. Reducing the uncertainty in the experimental energy levels, in particular a re-determination of the $n=13$ levels, 
 would be desirable both to clarify the role of the $4\mathrm{d}n\mathrm{f}$ channel and to improve
the models.

\begin{figure}[tb]
\centering
\includegraphics{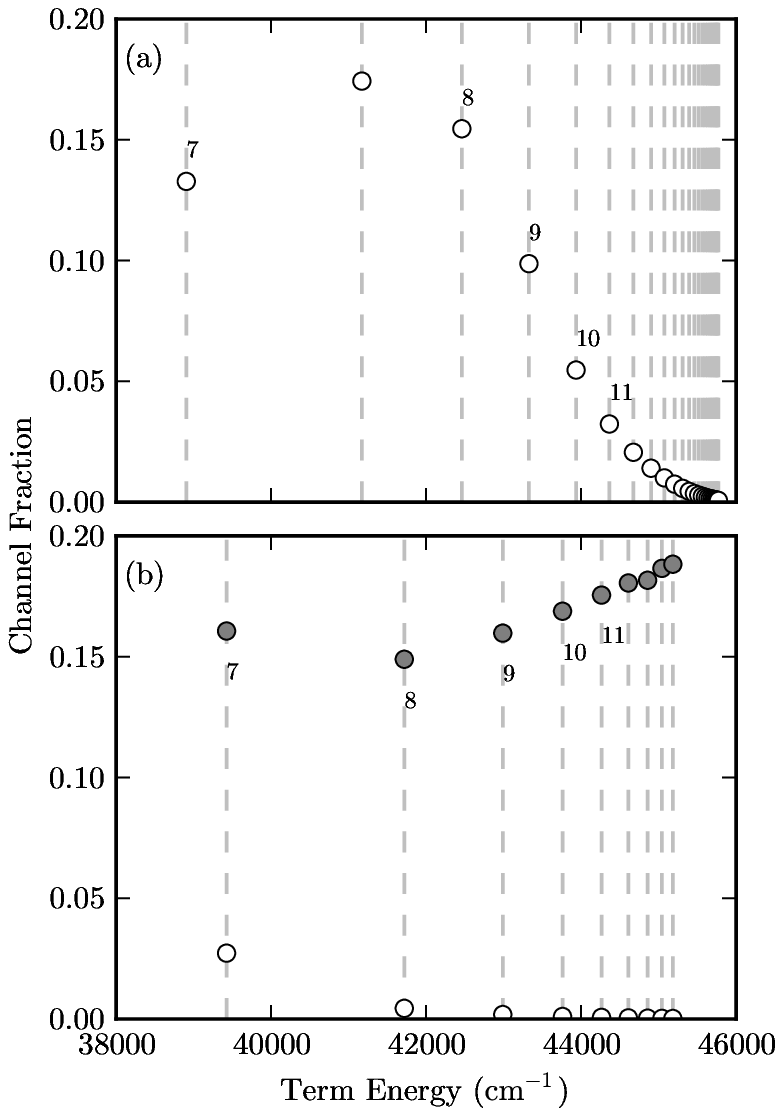}
\caption{Channel fractions of the perturbing $4\mathrm{d} n\mathrm{p}$ (white circles) and $4\mathrm{d}n\mathrm{f}$ (grey circles) channels in the $5\mathrm{s}n\mathrm{p} \; ^1\mathrm{P}_1$  and $^3\mathrm{P}_1$  series of strontium. Panel (a): $^1\mathrm{P}_1$ series ($7\le n \le 29$). Panel (b): $^3\mathrm{P}_1$ series ($7\le n \le 15$). The vertical dashed lines indicate the experimental positions of the 5s$n$p energies and (in panel (a)) that of the 4d5p $^1\mathrm{P}_1$ energy.}
\label{pstatefraction}
\end{figure}

Channel fractions for $J=1$ are presented in panel (b) of figure \ref{pstatefraction}. We assign the 4d$nl$ channel that peaks at the 4d5p $^3$P$_1$ position (at  37\, 302.731 cm$^{-1}$ \cite{Sansonetti2010}) to the 4d$n$p configuration. We thus assign the other 4d$nl$ channel to the 4d$n$f configuration. Interestingly, the latter is significant and spreads out fairly evenly throughout the triplet P$^{\rm   }$ states.
 
\begin{figure}[tb]
\centering
\includegraphics{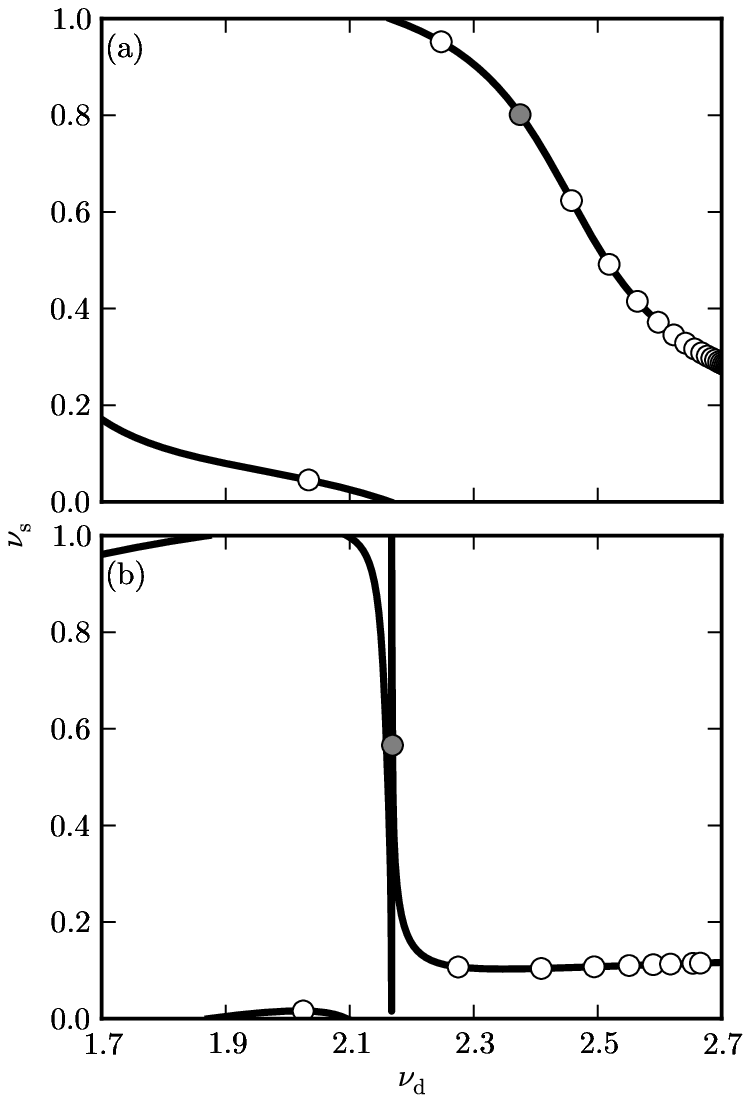}
\caption{Lu-Fano plot for the $5\mathrm{s}n\mathrm{p} \; ^1\mathrm{P}_1^{\rm   }$ series (panel (a)) and the $5\mathrm{s}n\mathrm{p} \; ^3\mathrm{P}_1^{\rm   }$ series (panel (b)). The $5\mathrm{s}n\mathrm{p} \; ^3\mathrm{P}_0^{\rm   }$ and $^3\mathrm{P}_2^{\rm   }$ states are similar to the 5s$n$p $^3\mathrm{P}_1^{\rm   }$ states. The open circles represent the experimental positions of the 5s$n$p bound states, the grey circles those of the 4d5p $^{1}\mathrm{P}_1^{\rm   }$ and $^{3}\mathrm{P}_1^{\rm   }$ perturbers. $\nu_\mathrm{s}$ is calculated relative to $I_\mathrm{s}= 45 \, 932.1982 \; \mathrm{cm}^{-1}$ and $\nu_\mathrm{d}$ relative to the average of the fine structure components of the $4\mathrm{d}$ ionization limits, $I_\mathrm{d}= 60 \, 628.26 \; \mathrm{cm}^{-1}$. The narrow resonance with the triplet perturber visible in panel (b) occurs between $n=6$ and $n=7$.} 
\label{pstatelufano}
\end{figure}

The Lu-Fano plot for the $^3$P$_1$ series shown in panel (b) of
figure~\ref{pstatelufano} appears to be very similar to that presented for triplet P-states in \cite{Aymar1987}; however the plot of figure~\ref{pstatelufano} refers specifically to the $J=1$ states whereas the fine structure of the P states was not resolved in \cite{Aymar1987}. 

\subsection{{$5\mathrm{s}n\mathrm{d}\; ^{1}\mathrm{D}_2$ and 
$^{3}\mathrm{D}_2$ states}}
The $^{1}\mathrm{D}_2$ and $^{3}\mathrm{D}_2$ series are particularly complex 
in being strongly coupled to each other. 
A previous empirical 
MQDT study \cite{Esherick1977} has shown that this coupling goes so far as to cause
an interchange in the
character of these states around 
$n=15$, making an accurate description of these series difficult. 
The MQDT model which was used in this previous work 
includes five channels ($5\mathrm{s}n\mathrm{d} \; ^1\mathrm{D}_2$,
$5\mathrm{s}n\mathrm{d} \; ^3\mathrm{D}_2$, $4\mathrm{d}n\mathrm{s} \; ^1\mathrm{D}_2$, 
$4\mathrm{d}n\mathrm{s} \; ^3\mathrm{D}_2$ and $5\mathrm{p}n\mathrm{p} \; ^1\mathrm{D}_2$), 
but does not reproduce all the experimental energy levels to within their error.
To the best of our knowledge, only one \emph{ab initio} investigation of this 
series has been carried out \cite{Aymar1987}. 
As spin-orbit coupling was not included in this work, however, 
there is as yet no \emph{ab initio} calculation of singlet-triplet mixing 
in these series. Empirical MQDT models of the D states of strontium
neglecting singlet-triplet mixing have also been proposed, including
only three channels for the singlet series and two channels for the triplet series;
however these models fail to reproduce the experimental 
data to within error \cite{Dai1995b}.

More recent measurements of the energy levels of the singlet D states \cite{Beigang1982a}
have reduced the experimental uncertainties
by a least an order of magnitude compared to the data used in the five-channel MQDT model of
 \cite{Esherick1977}, which makes it useful to revisit these two series. However, as 
the triplet states 
have not yet been measured more accurately, there is now a significant discrepancy 
in the precision
to which the two sets of 
energy levels are known. This discrepancy increases the complexity of the calculation, 
as the fitting algorithm favours the fit to the singlet energy levels, 
owing to their smaller uncertainty,
and gives less weight to the triplet states.

We use a six-channel model which includes the $4\mathrm{d}n\mathrm{d}$
channel
in addition to all the five channels of the model
of  \cite{Esherick1977}. This sixth channel is introduced to describe the perturbation of the 
$5\mathrm{s}n\mathrm{d}$ D$_2$ series by the 
$4\mathrm{d}^2 \; ^3\mathrm{P}_2$-even state around $n= 12$. This state 
has been measured at 
44\,~729.627 cm$^{-1}$ \cite{Sansonetti2010}.

Overall, the six channel model fits to 
$\chi^2_\nu=11.1$, including all the 5s$n$d energy levels for $7\le n\le 30$ and the $4\mathrm{d}^2 \; ^3\mathrm{P}_2$-even
perturber, 
with the deviations from experiment randomly scattered throughout the series. 
The reason for the deviations 
is unclear; they may be due to the poor knowledge of the triplet state energies. 
Fitting the six-channel model to the experimental data from \cite{Esherick1977},
which have much larger error bars, results 
in $\chi^2_\nu=1.4$; for comparison, the model presented in \cite{Esherick1977} fits with $\chi^2_\nu \sim 36$.

\begin{figure*}[tb]
\centering
\includegraphics{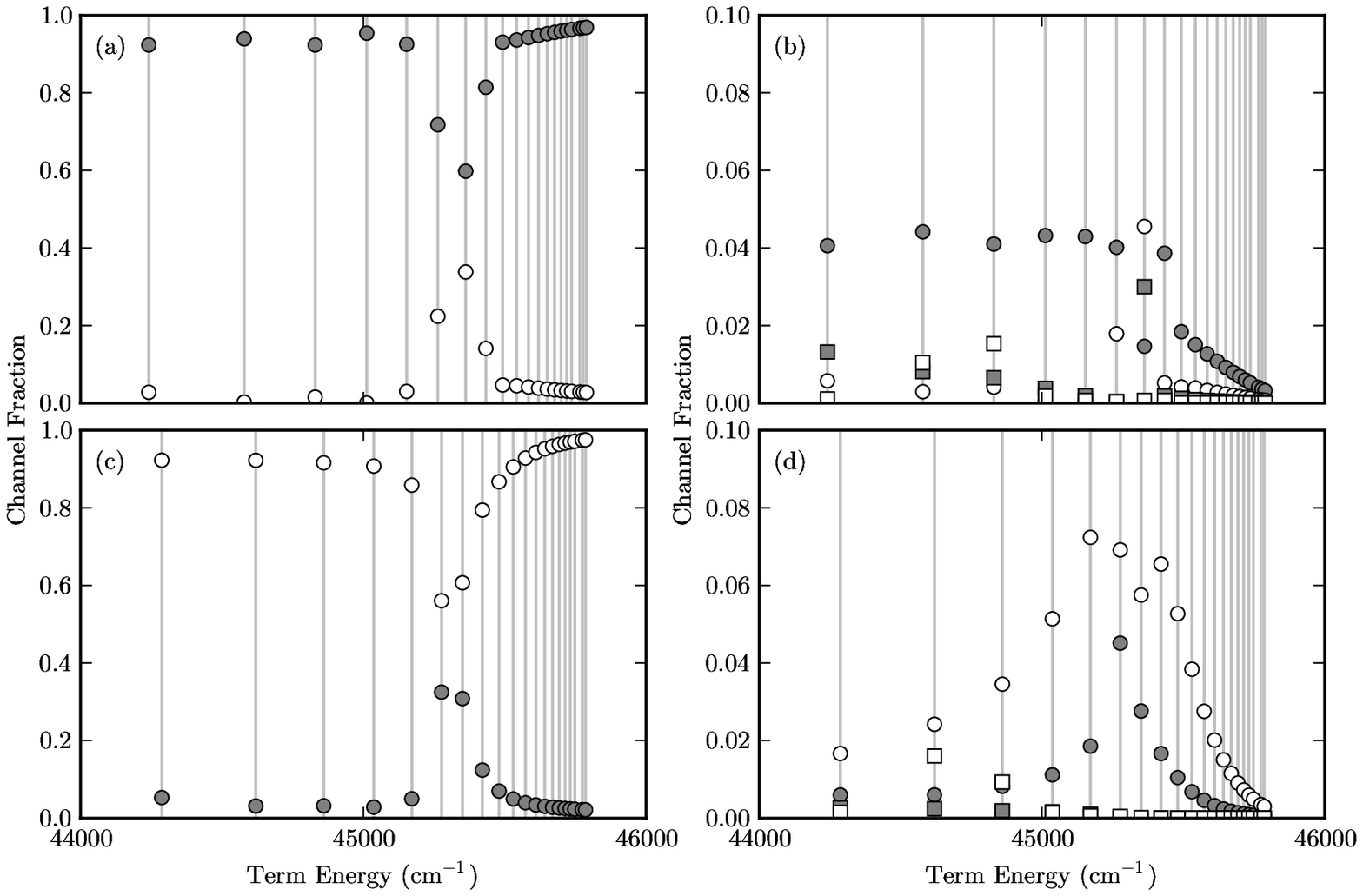}
\caption{Channel fractions in the $5\mathrm{s}n\mathrm{d} \; ^{1}\mathrm{D}_2$ and the $5\mathrm{s}n\mathrm{d} \; ^{3}\mathrm{D}_2$ series of strontium ($10\le n \le30$). The vertical lines indicate the experimental positions of the 5s$n$d bound state energies. Panels (a) and (b): $^1\mathrm{D}_2$ series (no data is available for the 5s27d $^1\mathrm{D}_2$ energy). Panels (c) and (d): $^3\mathrm{D}_2$ series (no data is available for the 5s28d $^3\mathrm{D}_2$ energy). (Following \cite{Esherick1977}, the singlet and triplet labels assigned to these states are interchanged between $n=15$ and $n=16$). Panels (a) and (c): the $5\mathrm{s}n\mathrm{d} \; ^{1}\mathrm{D}_2$ (grey circles) and $5\mathrm{s}n\mathrm{d} \; ^{3}\mathrm{D}_2$ channels (white circles). Panels(b) and (d): the $4\mathrm{d}n\mathrm{s}\; ^1\mathrm{D}_2$ (grey circles), $4\mathrm{d}n\mathrm{s}\; ^3\mathrm{D}_2$ (white circles), $5\mathrm{p}n\mathrm{p}$ (grey squares, multiplied by 10) and $4\mathrm{d}n\mathrm{d}$ channels (white squares).}
\label{evendstatefraction}
\end{figure*}

The resulting channel fractions are presented in figure \ref{evendstatefraction}. They  clearly show that the two $5\mathrm{s}n\mathrm{d}$ series swap 
singlet and triplet character 
between $n = 15$ and $n=16$ \cite{Esherick1977}. 
The swap has been attributed in previous investigations to the interaction of these channels with (unobserved) $4\mathrm{d}6\mathrm{s} \; ^1\mathrm{D}_2$ and $4\mathrm{d}6\mathrm{s} \; ^3\mathrm{D}_2$ states \cite{Esherick1977,Aymar1987}. The increase in the corresponding channel fractions around $n=15$ visible in the right-hand side of figure \ref{evendstatefraction} is consistent with this interpretation. The figure also shows that while the admixture of the perturbers to the $5\mathrm{s}n\mathrm{d}$ states decreases rapidly when $n$ increases above about 17, it is non-zero still for the highest states considered. This admixture could thus remain significant in high Rydberg states for observables that would depend sensitively on these perturbers. The channel fractions of the 4d$n$s and 5p$n$p channels in the 5s$n$d $^1\mathrm{D}_2$ and $^3\mathrm{D}_2$ states scales approximately like $n^{-3}$ at high $n$.

The $4\mathrm{d}^2 \; ^3\mathrm{P}_2$-even perturber is also described by this model, which shows that 
this state is $99\%$ $4\mathrm{d}^2$ in character. The admixture of
the 4d$n$d channel in the 
surrounding singlet and triplet 5s$n$d states, while small, is visible in figure \ref{evendstatefraction}.

The fit does not include the bound state conventionally labelled as the 5p$^2$ $^1\mathrm{D}_2$ state, at 36\,~960.842 cm$^{-1}$ \cite{Sansonetti2010}, as it lies too low in energy. MCHF calculations and
eigenchannel $R$-Matrix calculations \cite{Aymar1987,Aspect1984,Vaeck1988} found that this state has
predominantly a 4d$^2$ character, which is also supported by the analysis of its isotope shift
\cite{Aspect1984}. (A previous MCHF calculation \cite{FroeseFischer1981} had, however, found the 
5p$^2$ configuration to be predominant in this state.) We note, in this respect, that in our MQDT model 
the admixture of the 5p$n$p channel in the 5s$n$d $^1\mathrm{D}_2$ states increases as the energy decreases, without concomitant increase in the admixture of the 4d$n$d channel (panel (b) of figure \ref{evendstatefraction}). However, as the corresponding channel fractions are small, it would be imprudent to draw conclusions from these trends. The results of section \ref{sectionsingletS}
indicate that replacing our single 4d$n$d channel by channels converging to the fine-structure
resolved 4d$_{3/2}$ and 4d$_{5/2}$ ionization limits could be useful for a finer analysis of the
contribution of this perturber; however, the resulting large number of coupled channels would make
an empirical MQDT analysis prohibitive.  

\begin{figure}[tb]
\centering
\includegraphics{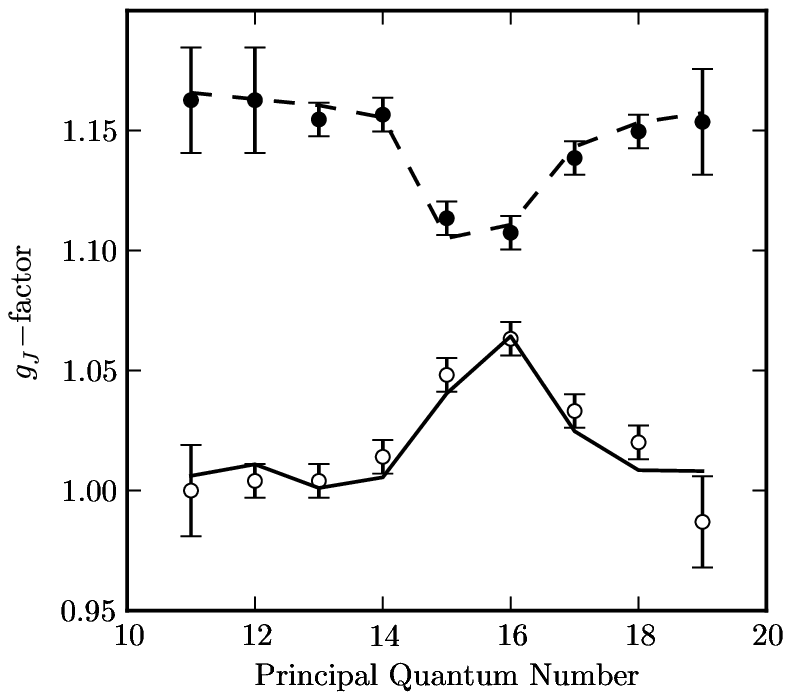}
\caption{Land{\'e} $g_J$-factors for the $\mathrm{D}_2$ states of strontium. The solid line represents the MQDT results for the states usually referred to as singlet states for $n < 16$ and triplet states for $n \geq 16$, and the dashed line those for the states usually referred to as triplet states for $n < 16$ and singlet states for $n \geq 16$. The experimental values are from \cite{Wynne1977} and have been re-normalised so that $g_J = 1$ for the singlet state at $n=11$ (the data presented in \cite{Wynne1977} were normalised so that $g_J = 1$ for the singlet state at $n=12$, which does not seem optimal since this state is more significantly perturbed by the $4\mathrm{d}^2 \; ^3\mathrm{P}_2$ state).}
\label{dstategfactors}
\end{figure}

Land\'e $g_J$-factors have previously been used in the fitting to improve empirical models 
for barium \cite{Aymar1984}. In the $\mathrm{D}_2$ states of strontium, however,
assigning particular
values of the $L$, $S$ and $J$ quantum numbers to the different channels
 is difficult around $n=15$ and 16, which makes the use
of Land\'e $g_J$-factors in the determination of
 the parameters of the model problematic.
Instead, the channel quantum numbers can be determined once the optimum fit to the energy 
levels is 
obtained, and the $g_J$-factors can then be calculated 
with the knowledge of the channel fractions 
and the symmetries of the states involved. For $LS$-coupled atoms, the Land{\'e} $g_J$-factor 
of a state is given by \cite{BransdenJoachain2003}
\begin{equation}
g_J = 1 + \frac{J(J+1) + S(S+1)- L(L+1)}{2J(J+1)}.
\label{lande}
\end{equation}
In the case where there is a breakdown of $LS$-coupling, such as in the $\mathrm{D}_2$ states 
of strontium, MQDT channel fractions can be used to find the total $g_J$-factor as a
weighted sum of 
individual channel-$g_J$-factors. The total $g_J$-factor is given by \cite{Wynne1977}
\begin{equation}
g_J = \sum_i {\bar A}_i^2 g_J^{(i)},
\label{mqdtgfactor}
\end{equation}
where the $g_J^{(i)}$-factors pertain to individual $LS$-coupled channels and are 1 for
$^1\mathrm{D}_2$ channels, 
$7/6$ for $^3\mathrm{D}_2$ channels and $3/2$ for $^3\mathrm{P}_2$ channels. 
The resulting $g_J$-factors are shown 
in figure \ref{dstategfactors}. The predictions of the present MQDT model
are seen to be in excellent agreement with experiment \cite{Wynne1977}.

\subsection{$5\mathrm{s}n\mathrm{d}\; ^{3}\mathrm{D}_1$ and $^{3}\mathrm{D}_3$ states}
\label{odddstates}
The $^3\mathrm{D}_1$ and $^3\mathrm{D}_3$ states are studied in \cite{Beigang1983}, which
provides 
experimental data and two-channel empirical models. However, for many states 
the calculated energies differ from the experimental values by
several standard deviations.
\cite{Aymar1987} also investigated the triplet D states, 
using an {\it ab initio} two-channel model, however without resolving the fine structure components.

We use three channels to fit the $^3\mathrm{D}_1$ and $^3\mathrm{D}_3$ states,
namely the 5s$n$d channel and two 4d$nl$ channels (the 4d$n$s channel had
already been recognized as significant by previous investigators
\cite{Beigang1983,Aymar1987}, and to this we add the 4d$n$d channel
in view of its importance in the $^{1}\mathrm{D}_2$ and $^{3}\mathrm{D}_2$ states). The resulting models 
reproduce the experimental energy levels 
with $\chi^2_\nu= 26.1$ and $\chi^2_\nu= 2.0$ for the $^3\mathrm{D}_1$ and $^3\mathrm{D}_3$ 
states respectively. However, the $5\mathrm{s}16\mathrm{d} \; ^3\mathrm{D}_1$ 
and the $5\mathrm{s}22\mathrm{d} \; ^3\mathrm{D}_3$ experimental energy levels
are significantly discrepant. The quality of the fit for the $^3\mathrm{D}_1$ series 
improves to $\chi^2_\nu= 0.5$ 
when the $n=16$ data point is neglected, and that for the $^3\mathrm{D}_3$ series 
to $\chi^2_\nu= 0.6$ when 
the $n=22$ data point is neglected. 
The deviation of the energy levels of these two states from the rest of the data 
is unexplained.

\begin{figure}[tb]
\centering
\includegraphics{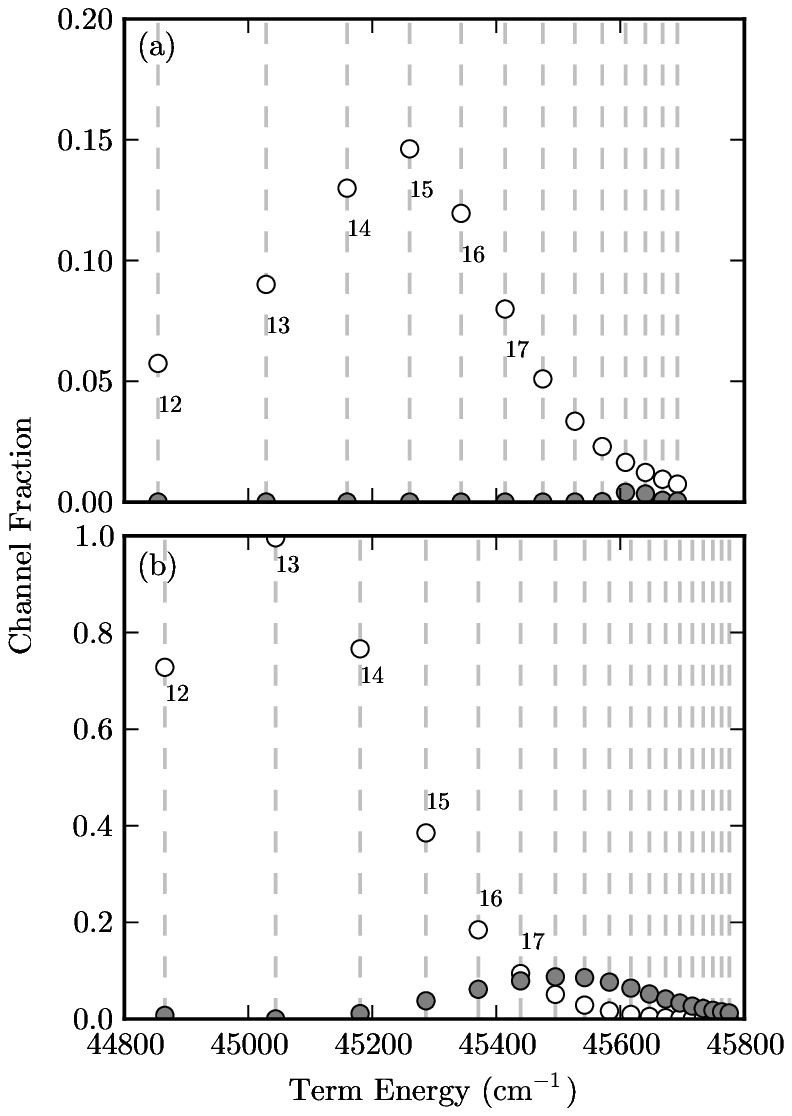}
\caption{Channel fractions in the $5\mathrm{s}n\mathrm{d} \; ^{3}\mathrm{D}_{1}$ and $5\mathrm{s}n\mathrm{d} \; ^{3}\mathrm{D}_{3}$ series of strontium.  Panel (a): $^{3}\mathrm{D}_{1}$ series ($12 \le n \le 24$). Panel (b): $^{3}\mathrm{D}_{3}$ series ($12 \le n \le 29$). White circles: $4\mathrm{d}n\mathrm{s}$ channel. Grey circles: $4\mathrm{d}n\mathrm{d}$ channel. The vertical dashed lines indicate the experimental positions of the 5s$n$d bound state energies.}
\label{odddstatefraction}
\end{figure}

The corresponding channel fractions show a large importance of one of the two 4d$nl$ channels, particularly in the $^3\mathrm{D}_{3}$ states (figure  \ref{odddstatefraction}). Given that the 4d$n$s channel had been found previously to be significant in the $^3\mathrm{D}$ states \cite{Beigang1983,Aymar1987}, we assign the channel making the largest contribution to this configuration, and thus assign the other channel to the 4d$n$d configuration. The magnitude and position of the perturbation also suggests that the perturbers are unobserved 4d6s  $^3\mathrm{D}_{1}$ and  $^3\mathrm{D}_{3}$ states rather than unobserved 4d$^2$ states of other symmetries. Remarkably, the state usually labelled 5s13d $^3\mathrm{D}_{3}$ is found to be almost 100\% in a 4d6s configuration. The admixture of the other 4d$nl$ channel is also found to be significant around 45\, 500 cm$^{-1}$ in the $^3\mathrm{D}_{3}$ states. However, these results should be considered as merely indicative, given the quality of the spectroscopic data they are based on. It would be useful to re-determine the $J=1$ and $J=3$ energy levels, and also compare the present empirical MQDT analysis to fine-structure-resolved {\it ab initio} calculations.

\subsection{{$5\mathrm{s}n\mathrm{f}\; ^{1}\mathrm{F}_3^{\rm  o}$  states}}
The F states of strontium have been less studied than the S, P and D states. 
As far as we know, the only MQDT models published so far are {\it ab initio} and only
address the $J=3$
series \cite{Kompitsas1990}. 
This previous work showed that the 5s$n$f  $^1\mathrm{F}_3^{\rm   }$ series is perturbed
by a  $4\mathrm{d}5\mathrm{p}$ $^1\mathrm{F}_3^{\rm   }$ perturber, and similarly that
the  5s$n$f  $^3\mathrm{F}_3^{\rm   }$ series is perturbed
by a  $4\mathrm{d}5\mathrm{p}$ $^3\mathrm{F}_3^{\rm   }$ perturber. 
A multiconfiguration Hartree-Fock calculation of the lowest $^1$F$_3$ states found an admixture of the 
$4\mathrm{d}5\mathrm{p}$ and $4\mathrm{d}4\mathrm{f}$ configurations
into the low-lying $5\mathrm{s}n\mathrm{f}$ states \cite{Vaeck1988,Hansen1977}.

We have carried out an empirical MQDT study of all the singlet and triplet 5s$n$f F series, using recently published experimental energies \cite{Sansonetti2010}.  We find the
 $^1\mathrm{F}_3^{\rm   }$ 
series to be very similar to the $^1\mathrm{P}_1^{\rm   }$ series, in that the data could be well fitted
by a two-channel model. Besides the 5s$n$f channel, the model includes the 4d$n$p channel in view
of the presence of a $4\mathrm{d}5\mathrm{p}$ $^1\mathrm{F}_3^{\rm   }$ state low in the spectrum
(the same configuration but a different energy and overall symmetry than the state perturbing
the   $^1\mathrm{P}_1^{\rm   }$ series). The experimental bound state energies are taken
from from \cite{Sansonetti2010} for $n$ up to 20; as the data reported in \cite{Sansonetti2010}  
do not extend higher in energy, we have used the less precise values of \cite{Rubbmark1978}
for $21\le n \le 29$. The resulting MQDT model fits the whole singlet series (from $n=4$ upwards)
as well as the perturber with $\chi_\nu^2=0.8$.

\begin{figure}[tb]
\centering
\includegraphics{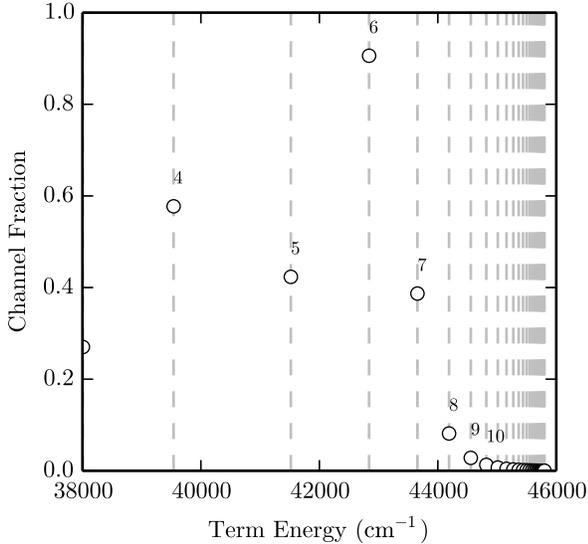}
\caption{Channel fraction of the perturbing $4\mathrm{d}n\mathrm{p}$ channel in the $^1\mathrm{F}_3^{\rm   }$ series of strontium. The vertical dashed lines indicate the experimental positions of the bound state energies (the data point at 38\,~007.742 cm$^{-1}$ is the $4\mathrm{d}5\mathrm{p}$ $^1\mathrm{F}_3^{\rm   }$ state).}
\label{singletfstatefraction}
\end{figure}

As pointed out by previous investigators \cite{Vaeck1988,Kompitsas1990,Hansen1977}, and
as is illustrated by figure \ref{singletfstatefraction}, the $4\mathrm{d}5\mathrm{p}$ $^1\mathrm{F}_3^{\rm   }$ state (at
38\,~007.742 cm$^{-1}$ \cite{Sansonetti2010}) strongly perturbs the low end of 
the $^1\mathrm{F}_3^{\rm   }$ series. In our model, the admixture of
the 4d$n$p channel is particularly large in the 5s6f state. In the 5s4f state, it somewhat exceeds the admixture of 37\% 
found in {\it ab initio} calculations, whereas it is somewhat less than
the admixture of 50\% these calculations found for the $4\mathrm{d}5\mathrm{p}$ $^1\mathrm{F}_3^{\rm   }$ \cite{Kompitsas1990}. These differences between the
present empirical model and the {\it ab initio} model have no obvious reasons; they
may point to the limitations of an MQDT analysis for states so low in the spectrum.

\subsection{{$5\mathrm{s}n\mathrm{f}\; ^{3}\mathrm{F}_2^{\rm  o}$, $^{3}\mathrm{F}_3^{\rm  o}$ and $^{3}\mathrm{F}_4^{\rm  o}$ states}}

The experimental $^3$F energy levels of \cite{Moore1952} have been superseded by the more precise results of \cite{Rubbmark1978} and \cite{Sansonetti2010}, the latter giving fine-structure resolved energy levels up to the 5s20f $^3$F states. These data are sufficiently extensive to be taken as a basis for an empirical MQDT analysis of these series complementing the eigenchannel $R$-matrix  calculations of \cite{Kompitsas1990}. This previous work, which only addressed the $J=3$ states, pointed to the advantage of using $jj$-coupled channels rather than $LS$-coupled channels in analysing these series. For the $^3$F$_3$ states, this {\it ab initio} MQDT analysis was based on five channels, namely the 5s$_{1/2}n$f$_{5/2}$, 5s$_{1/2}n$f$_{7/2}$, 4d$_{3/2}n$p$_{3/2}$, 4d$_{5/2}n$p$_{1/2}$ and 4d$_{5/2}n$p$_{3/2}$ channels. However, in our work we treat the three series in $LS$-coupling. Indeed, only taking into account $jj$-coupled 5s$n$f and 4d$n$p channels would mean that for the $J=4$ states the MQDT model would include only the  5s$_{1/2}n$f$_{7/2}$ and 4d$_{5/2}n$p$_{3/2}$ channels, and as we have found, such a two-channel model is unsatisfactory (we could not find MQDT parameters for which the model fits the data well and in the same time does not yield an excessively large 4d$n$p channel fraction for all the low $n$ states). Adding the 4d$n$f configuration to this model, to improve the fit, would require the inclusion of an excessively large number of channels in a $jj$-coupled scheme. We therefore prefer to work in $LS$-coupling. Given that the three  $^3$F series are very similar to each other in terms of quantum defects, we use the same choice of channels for all three, namely one channel converging to the 5s$_{1/2}$ ionization limit and two channels converging to the average of the 4d$_{3/2}$ and 4d$_{5/2}$ ionization limits. This scheme is the  same as for the $^3$P states (see section \ref{pstates}), but here, of course, the three channels should be assigned to 5s$n$f $^3$F, 4d$n$p $^3$F and 4d$n$f $^3$F symmetries.

\begin{figure}[tb]
\centering
\includegraphics{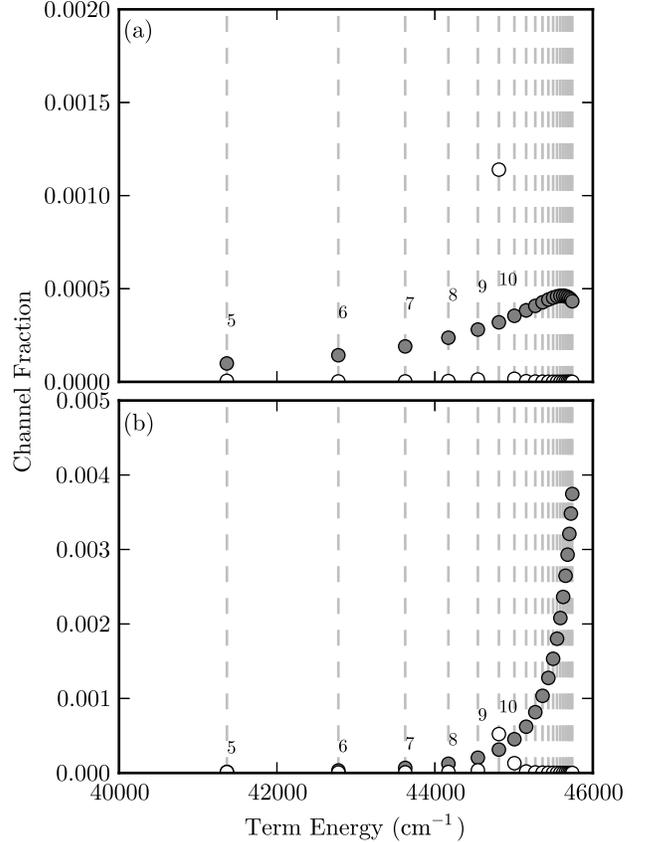}
\caption{Channel fractions of the perturbing $4\mathrm{d}nl$ (white circles) and $4\mathrm{d}nl'$ (grey circles) channels in the $^3\mathrm{F}_3$ and $^3\mathrm{F}_4$ series of strontium. Panel (a): $^3\mathrm{F}_3$ series. Panel (b): $^3\mathrm{F}_4$ series. The vertical dashed lines indicate the experimental positions of the bound state energies. As discussed in the text, it is likely that the white circles correspond to states of 4d$n$p symmetry and the grey circles to states of 4d$n$f symmetry.}
\label{tripletfstatefractions}
\end{figure}

\begin{figure}[tb]
\centering
\includegraphics{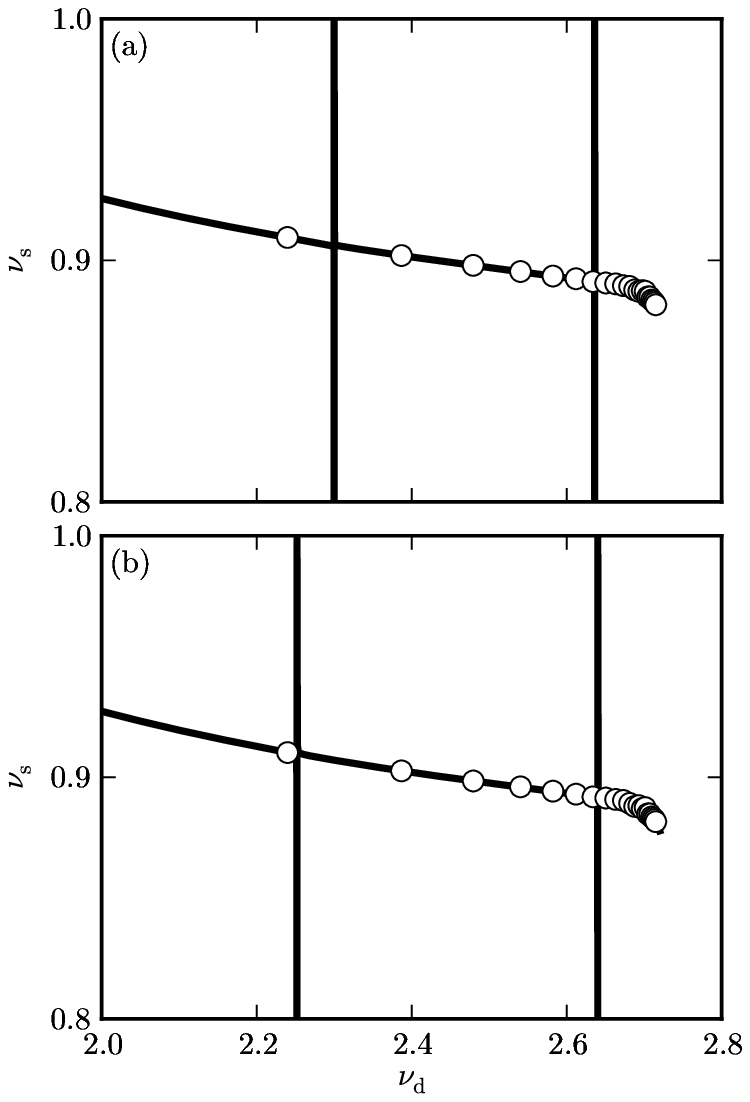}
\caption{Lu-Fano plot for the $5\mathrm{s}n\mathrm{f} \; ^3\mathrm{F}_3^{\rm   }$ series (panel (a)) and the $5\mathrm{s}n\mathrm{f} \; ^3\mathrm{F}_4^{\rm   }$ series (panel (b)). The $5\mathrm{s}n\mathrm{f} \; ^3\mathrm{F}_2^{\rm   }$ are similar to the 5s$n$f $^3\mathrm{F}_3^{\rm   }$ states. The open circles represent the experimental positions of the 5s$n$f bound states. The vertical lines are narrow resonance structures. $\nu_\mathrm{s}$ is calculated relative to $I_\mathrm{s}= 45 \, 932.1982 \; \mathrm{cm}^{-1}$ and $\nu_\mathrm{d}$ relative to the average of the fine structure components of the $4\mathrm{d}$ ionization limits, $I_\mathrm{d}= 60 \, 628.26 \; \mathrm{cm}^{-1}$. } \label{tripletfstatelufano}
\end{figure}

We fit the parameters of the models to the energy levels of \cite{Sansonetti2010} for $4 \leq n \leq 20$ and to those of \cite{Rubbmark1978} for $21 \leq n \leq 24$ ($\chi^2_\nu = 0.6$, 0.7 and 0.3 for, respectively, the $^3\mathrm{F}_2^{\rm   }$, $^3\mathrm{F}_3^{\rm   }$ and $^3\mathrm{F}_4^{\rm   }$ series). The resulting channel fractions are shown in figure~\ref{tripletfstatefractions}, and Lu-Fano plots can be found in figure~\ref{tripletfstatelufano}. We identify the channel represented by white circles with the 4d$n$p channel in view of its much larger admixture in the 5s4f $^3\mathrm{F}_J^{\rm   }$ states (not represented in figure~\ref{tripletfstatefractions}), which is consistent with their expected interaction with the 4d5p $^3\mathrm{F}_J^{\rm   }$ perturbers \cite{Kompitsas1990}. The grey circles thus refer to the 4d$n$f channel. No results are shown for the $^3\mathrm{F}_2^{\rm   }$ states as they are almost identical to those for the $^3\mathrm{F}_3^{\rm   }$ states.

As seen from these two figures, the triplet F bound states are largely unperturbed over the whole series. Narrow structures can be seen in both panels of figure \ref{tripletfstatelufano}, which correspond to the peaks in the white circles in \ref{tripletfstatefractions}. These features are not observed in the \emph{ab initio} models of \cite{Kompitsas1990}, and remain unexplained; the features may be due to the large uncertainties in the energy level data, however the resonances in the Lu-Fano plot do not change significantly when adding random perturbations of the order of the uncertainties to the data. We note that no doubly excited bound states have been measured near these energies.

The Lu-Fano plots in figure \ref{tripletfstatelufano} also show the beginning of a curve near $\nu_\mathrm{d} \sim 2.7$, which corresponds to the perturbation described by the grey circles in figure \ref{tripletfstatefractions}. The trend in the experimental values of $\nu_\mathrm{s}$ are smooth and consistent with a previously neglected perturbation. A deviation in the experimental data from stray electric and magnetic fields, however, cannot be neglected; if the trend in the values of $\nu_\mathrm{s}$ are caused by stray fields, then the $4\mathrm{d}n\mathrm{f}$ channel may not correspond to the presence of a perturber, but may in fact be describing the influence of external fields. The issue of the presence of a $4\mathrm{d}4\mathrm{f}$ perturber can only be addressed with improved experimental values of the energy levels.

\section{Radiative lifetimes}
\label{lifetimesection}
MQDT can be used to calculate natural radiative lifetimes. This calculation is most commonly done by assuming that the radiative width of the states varies with the effective principal quantum number $\nu_i$ according to the equation
\begin{equation}
\Gamma(\nu_i) = \sum_i \frac{{\bar A}_i^2 \Gamma_i}{\nu_i^3},
\label{channelwidths}
\end{equation}
where $\Gamma(\nu_i)$ is the radiative width. The constants $\Gamma_i$, the ``channel widths", are obtained by fitting this expression to the experimental lifetimes. This method was first used to study lifetimes in the heavily perturbed series of barium and has been applied to various other species since \cite{Aymar1984}. In particular, the method has been used to study the lifetime of  $^1\mathrm{S}_0$, $^1\mathrm{D}_2$ and $^3\mathrm{D}_2$ states of strontium \cite{Dai1995a,Dai1995b}. For the $^1\mathrm{S}_0$ states, excellent agreement was obtained with experiment for the states included in the fit; however, there were large differences for the lifetimes of the states higher in the series \cite{Grafstrom1983, Millen2011b}. The agreement with the experimental lifetimes  was excellent for the $^1\mathrm{D}_2$ and $^3\mathrm{D}_2$  states, too, although the MQDT models used in this work neglected single-triplet mixing (and could therefore not reproduce the measured energy levels well). Combined $R$-matrix and MQDT calculations have also been carried out for some of the lowest lying states of strontium \cite{Werij1992}. These \emph{ab initio} calculations give results generally in good agreement with experiment but do not go higher than $n=10$.

The method of fitting channel widths, however, does not provide a consistent description of lifetimes across different series. As each series is treated separately, the channel widths that are fitted to the experimental lifetimes can be made to fit regardless of the quality of the MQDT model used or the underlying configurations of the perturbers. The channel widths description thus cannot be extended beyond a purely phenomenological description of the radiative lifetimes of the Rydberg states.

In this work, we calculate the lifetimes of $^1\mathrm{S}_0$ and $^1\mathrm{D}_2$ states by combining the MQDT analysis outlined in the previous section with the calculation of dipole matrix elements within the Coulomb approximation \cite{Seaton2002}. As the dipole matrix elements for doubly excited states cannot be calculated within the Coulomb approximation, however, we leave the radial parts of these matrix elements as free parameters. In contrast with the method of the channel widths, the radial dipole matrix elements obtained from this fitting procedure are consistent across different series, and provide a means to extend the MQDT analysis to other calculations.

We start from the general expression for the spontaneous decay width for a transition from a state $a$ to a state $b$ \cite{BransdenJoachain2003}, namely
\begin{equation}
\Gamma_{ba} = \frac{4 \alpha}{3c^2} |\omega_{ba}|^3 | \langle \Psi_b | \vec{d} \,| \Psi_a \rangle |^2,
\label{generalwidth}
\end{equation}
where $\vec{d}$ is the dipole operator and $\omega_{ba}$ is the Bohr transition frequency between the two states. In the notation used here,  $\omega_{ba}= 2\pi c(E_a - E_b)$. The corresponding oscillator strength is given by the equation \cite{Hilborn1982}
\begin{equation}
f_{ab} = \frac{2J_b +1}{2J_a + 1} \frac{c^2  \Gamma_{ba}}{2 \alpha \omega_{ba}^2 }.
\label{oscstrength}
\end{equation}
Taking spontaneous decay only into account, the total radiative width of state $a$, $\Gamma_a$, is obtained by summing over all the states $b$ dipole-coupled to it:
\begin{equation}
\Gamma_a = \sum_b \Gamma_{ba}.
\label{eq:totalwidth}
\end{equation}
However this result needs to be corrected for the effect of the ambient blackbody radiation. Following \cite{Farley1981}, we write that at a temperature $T$
\begin{equation}
\begin{split}
\Gamma_a &= \sum_{b, E_b < E_a} \Gamma_{ba} \left(1 + \frac{1}{e^{|\omega_{ba}|/k_\mathrm{B}T} -1}\right)\\
&+ \sum_{b,E_b>E_a} \frac{\Gamma_{ba} }{e^{|\omega_{ba}|/k_\mathrm{B}T} -1},
\end{split}
\label{bbr}
\end{equation}
where the first sum runs over the states lower in energy than the state $a$ and
the second sum over the states higher in energy.  The contribution of the thermal photons decreases very rapidly when the difference $|n_a-n_b|$ between the principal quantum numbers of the initial and final states increases. For the states
considered in this paper, truncating the second sum to $n_b \le n_a+5$ is sufficient to ensure that $\Gamma_a$ has converged to three
significant figures. The temperature used in the calculation is that at which
the experimental lifetimes considered were measured, i.e., either 300 K or 77 K.

We work in $LS$-coupling throughout. With  $l_{1i}$ and $l_{2i}$ denoting the orbital angular momentum quantum number of, respectively, the inner electron and the Rydberg electron in channel $i$,
\begin{equation}
\begin{split}
\Gamma_{ba} = \frac{4 \alpha}{3c^2} &|\omega_{ba}|^3 \left[\sum_{i,j} 
(-1)^{l_{1ai} + l_{2,\mathrm{max}}+S_{ai}}
{\bar A}_{ai}{\bar A}_{bj} \right. \\ & \times  \sqrt{l_{2,\mathrm{max}} (2L_{bj} +1) (2J_{b}+1)(2L_{ai}+1)}\\
&\times \left.\begin{Bmatrix}
J_{b} & 1 & J_{a}\\
L_{ai} & S_{ai} & L_{bj}
\end{Bmatrix}^2 \begin{Bmatrix}
L_{bj} & 1 & L_{ai}\\
l_{2 ai} & l_{1 ai} & l_{2 bj}
\end{Bmatrix}^2 R_{bj,ai} \right]^2,
\end{split}
\label{width}
\end{equation}
where $R_{bj,ai}$ is a radial matrix element and $l_{2,\mathrm{max}} = \max (l_{2ai},l_{2bj})$ (see appendix \ref{appendixB}). As the channels are summed over coherently, the resulting dipole matrix elements include cross terms between channels. These terms can make a significant contribution to the lifetimes.

As we usually do not include states lying below $E\sim 38\,000 \; \mathrm{cm}^{-1}$, the correspondingly poor description of the lowest lying states in the MQDT models is an issue in this approach. The lifetimes and oscillator strengths considered here depend strongly on the $5\mathrm{s}^2\; ^1\mathrm{S}_0$, $5\mathrm{s}5\mathrm{p} \; ^1\mathrm{P}_1$ and $5\mathrm{s}4\mathrm{f} \; ^1\mathrm{F}_3$ states, which are not included in our MQDT models. We use the weights of the various configurations in the MCHF wave functions of \cite{Vaeck1988} in lieu of the MQDT mixing coefficients ${\bar A}_i$ to describe these low-lying states.

We calculate the radial matrix elements for the transitions between two different 5s$nl$ channels by assuming that the orbital of the inner electron is the same in the initial and final states and that the Rydberg electron is described by Coulomb functions \cite{Seaton2002}, as given by equation (\ref{longrange}). Hence
\begin{equation}
R_{bj,ai} = \int_0^\infty \phi_{bj} (r) \phi_{ai}(r) r dr
\label{radialmatrixelement}
\end{equation}
for these transitions. This one-electron approximation, however, is not expected to be suitable for transitions involving other channels, as those arise from the mixing of the 5s$nl$ states with low lying doubly excited states. Instead, we proceed as follows. We interpret the admixture with the 4d$n$s, 4d$n$p, 4d$n$d, 4d$n$f and 5p$n$p channels uncovered by the MQDT calculation as admixture with, respectively, the 4d6s, 4d5p, 4d$^2$, 4d4f and 5p$^2$ configurations, in agreement with the discussion above. Separating the widths $\Gamma_{ba}$ into radial integrals and angular factors then gives rise to the five radial integrals of Table \ref{radialmatrixelements}, which we treat as free parameters to be found by fitting the resulting widths and oscillator strenghts to the experimental data. To keep the number of fitting parameters low, we assume that the corresponding matrix elements depend only on the configuration of the initial and final states, not on their energy or overall symmetry.

The values of the $^1\mathrm{S}_0$ lifetimes used in the fit are obtained from \cite{Gornik1977,Grafstrom1983,Osherovich1979,Millen2011b} and those of the $^1\mathrm{D}_2$ state lifetimes from \cite{Bergstrom1986,Gornik1977, Grafstrom1983,Lochead2012}. In view of the unreliability of the MQDT models and the Coulomb approximation for low lying states we do not include states with $n_a < 10$ in the fit. However, besides these lifetimes
we also include the $5\mathrm{s}^2 \; ^1\mathrm{S}_0$ to $5\mathrm{s}n\mathrm{p} \; ^1\mathrm{P}_1$ and $5\mathrm{s}5\mathrm{p} \; ^1\mathrm{P}_1$ to $5\mathrm{s}n\mathrm{d} \; ^1\mathrm{D}_2$ oscillator strengths from \cite{Parkinson1976,Connerade1991,Mende1997,Mende1996} (the values in \cite{Mende1996} need to be divided by 3 to bring their values in line with the definition given above). Table \ref{radialmatrixelements} shows the values of the resulting radial dipole matrix elements involving doubly excited states. Estimates of the sensitivity of these five fitting parameters to the uncertainties in the experimental data are given in brackets. These estimates were obtained by giving random variations to the experimental values of the lifetimes and oscillator strengths used in the fit and reoptimizing the parameters.  It should be noted that the errors quoted do not reflect the fact that the results given in the Table \ref{radialmatrixelements} are model-dependent and may be affected by inaccuracies in the calculation of the matrix elements of the dipole operator between singly excited configurations.

\begin{table}
\begin{ruledtabular}
\begin{tabular}{cc}
Transition & Radial dipole matrix element\\
\hline
$\langle 4\mathrm{d}6\mathrm{s}| r | 4\mathrm{d}5\mathrm{p}\rangle$ & 6.48(5)\\
$\langle 5\mathrm{p}^2| r | 4\mathrm{d}5\mathrm{p}\rangle$ & -3.10(8)\\
$\langle 5\mathrm{p}^2| r | 5\mathrm{s}5\mathrm{p}\rangle$ & -32(2)\\
$\langle 4\mathrm{d}^2| r | 4\mathrm{d}5\mathrm{p}\rangle$ & -12.7(4)\\
$\langle 4\mathrm{d}^2| r | 4\mathrm{d}4\mathrm{f}\rangle$ & 28(3)\\
\end{tabular}
\end{ruledtabular}
\caption{Radial matrix elements involving doubly excited states, as obtained by fitting the calculated lifetimes and oscillator strengths to experimental values, in atomic
units.  An estimated error on the last digit quoted is shown in brackets, as discussed in the text.}
\label{radialmatrixelements}
\end{table}

\begin{figure}[tb]
\centering
\includegraphics{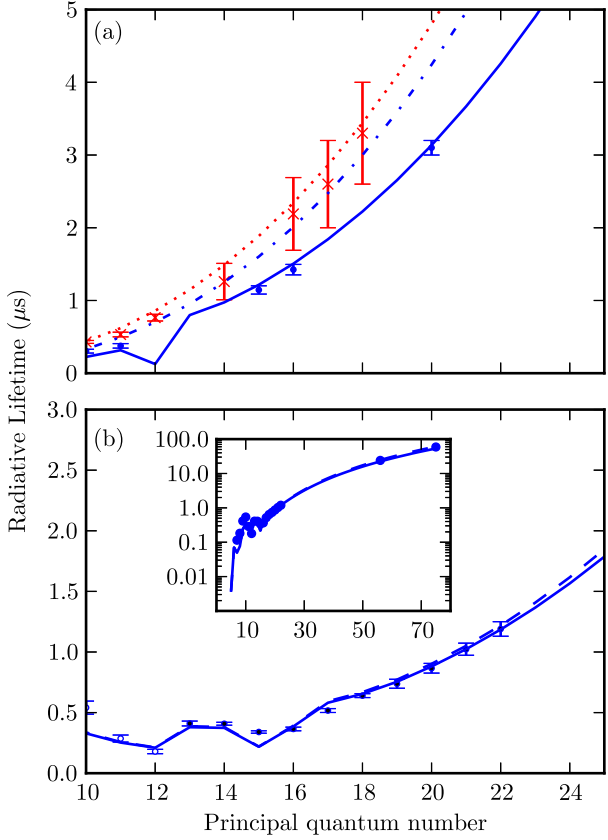}
\caption{(colour online) Panel (a): Lifetimes of the $^1\mathrm{S}_0$ states of strontium (blue circles, solid curve and dash-dotted curve) and of the $^2\mathrm{S}_{1/2}$ states of rubidium (red crosses and dotted curve) at 300 K \cite{Marek1980,Gounand1980,Theodosiou1984}. The solid curve represents the present MQDT results, the dotted and dash-dotted curves the results of a one-channel calculation in the Coulomb approximation, and the markers the experimental lifetimes for strontium $^1\mathrm{S}_0$ states \cite{Gornik1977,Grafstrom1983,Osherovich1979,Millen2011b}. Panel (b): Lifetimes of the $^1\mathrm{D}_2$ states of strontium, either at 300 K (blue filled circles and solid curve) or 77 K (blue open circles and dashed curve). The solid and dashed curves represent the present MQDT results and the markers the experimental lifetimes \cite{Gornik1977,Grafstrom1983,Bergstrom1986,Lochead2012}.}
\label{lifetimes}
\end{figure}

The resulting lifetimes are shown in figure ~\ref{lifetimes}. Generally, the trends across the $^1\mathrm{S}_0$ and $^1\mathrm{D}_2$ series are well predicted by the MQDT calculations, although the theory does not reproduce all the experimental data within their error ($\chi^2_\nu=3.8$ if the lifetime of the 5s15d $^1\mathrm{D}_2$ state is included in the calculation, or 1.57 if it is not). The oscillator strengths (not shown) are generally slightly overestimated; however, nearly all the calculated values fall within the error on the experimental values. It is clear from panel (a) of the figure that the Rydberg S states of strontium have a shorter lifetime than the S states of rubidium of a similar energy. The difference between the two species is even larger for the D states. It is mainly a consequence of the admixture of doubly excited states \cite{Bergstrom1986}: when this admixture is neglected and the S-series of strontium is described by a single-channel (5s$n$s) model, the calculated lifetimes are similar for the two species (albeit still slightly shorter for strontium --- compare the dash-dotted curve to the dotted curve in panel (a)). One can also note from the figure that for rubidium the Coulomb approximation gives lifetimes overall in good agreement with experiment. The fitting procedure is therefore not simply compensating for the errors from the use of Coulomb functions in the calculation, and the Coulomb approximation will contribute only a small uncertainty to the fitted radial matrix elements in table \ref{radialmatrixelements}.

\begin{figure}[tb]
\centering
\includegraphics{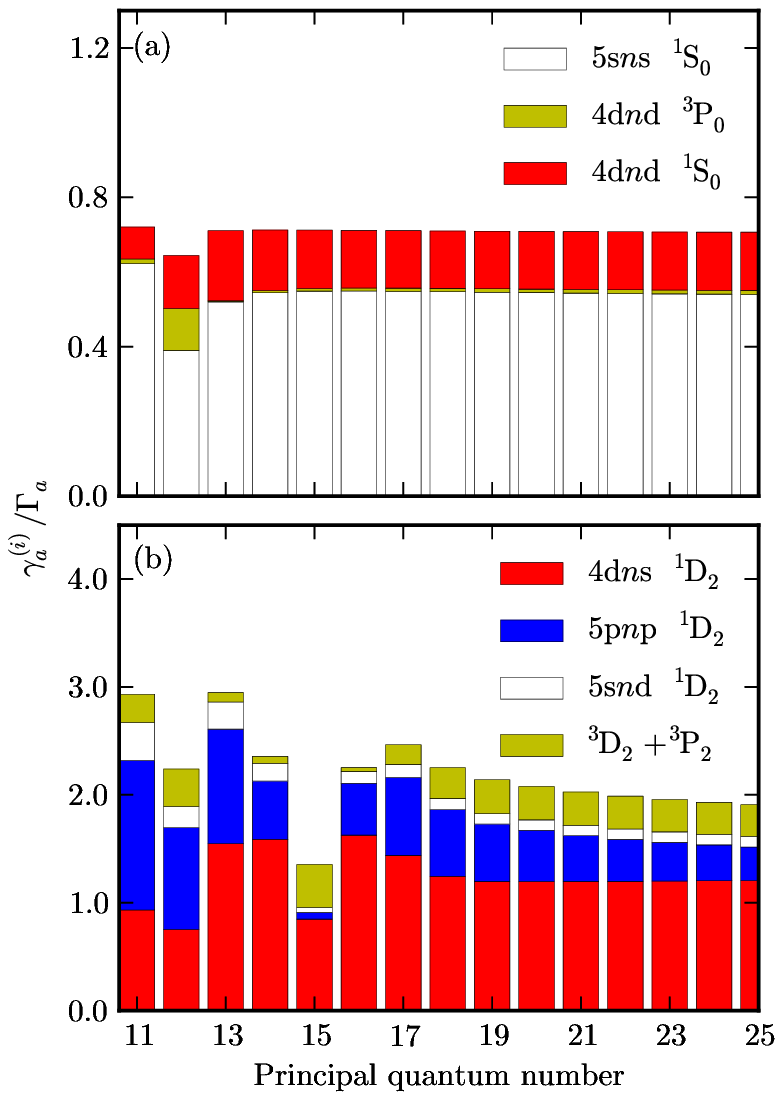}
\caption{(colour online) Normalized partial widths, as defined by \eqref{contributions}, for the $^1\mathrm{S}_0$ states (panel (a)) and $^1\mathrm{D}_2$ states (panel (b)). The colours label the different channels in each panel, as given by the corresponding keys. In panel (b), the $^3\mathrm{D}_2 + {^3\mathrm{P}_2}$ label denotes the sum of the $5\mathrm{s}n\mathrm{d} \; ^3\mathrm{D}_2$, $4\mathrm{d}n\mathrm{s} \; ^3\mathrm{D}_2$ and $4\mathrm{d}n\mathrm{d} \; ^3\mathrm{P}_2$ channels.}
\label{partialwidths}
\end{figure}

The impact of the perturbers on the lifetimes of the excited states of strontium
is also clear from figure \ref{partialwidths}. This figure shows how the normalized partial widths $\gamma_a^{(i)}/\Gamma_a$ vary with the principal quantum number
of state $a$. These
quantities are defined as follows:
\begin{equation}
\begin{split}
\frac{\gamma_a^{(i)}}{\Gamma_a}  & = 
\frac{1}{\Gamma_a} \frac{4 \alpha}{3c^2} {\bar A}_{ai}^2  \sum_{b}|\omega_{ba}|^3 \left[\sum_j 
(-1)^{l_{1ai} + l_{2,\mathrm{max}}+S_{ai}}
{\bar A}_{bj} \right. \\ & \times  \sqrt{l_{2,\mathrm{max}} (2L_{bj} +1) (2J_{b}+1)(2L_{ai}+1)}\\
&\times \left.\begin{Bmatrix}
J_{b} & 1 & J_{a}\\
L_{ai} & S_{ai} & L_{bj}
\end{Bmatrix}^2 \begin{Bmatrix}
L_{bj} & 1 & L_{ai}\\
l_{2 ai} & l_{1 ai} & l_{2 bj}
\end{Bmatrix}^2 R_{bj,ai} \right]^2,
\end{split}
\label{contributions}
\end{equation}
with $\Gamma_a$ calculated according to equation (\ref{eq:totalwidth}) (thus ignoring black body radiation). These partial widths indicate the importance in
the decay width 
of the different channels contributing to the initial state; however, they
ignore the interference between channels (there are no cross terms between different
values of $i$), and therefore they do not sum to 1. As seen from the figure, the perturbers are significant in the S-states and dominate  the lifetimes of the D-states, even for high principal quantum numbers. Indeed, both the radiative
lifetimes and 
the channel fractions of the perturbing 4d$n$s and
5p$n$p channels in the 5s$n$d $^1\mathrm{D}_2$ states
scale approximately like $n^{-3}$ at high $n$, which means that the relative contribution of
the doubly-excited component of the Rydberg wave functions to the decay width
tends to remain constant. As seen from panel (a) of the figure, the same also
applies to the Rydberg S states, but to a smaller scale. The admixture of the 4d6s perturber to
the D states over a range of principal quantum numbers reduces the 5s$n$d
lifetimes by about a factor of 3 compared to the lifetimes of the S states.

Besides these general trends, we also note that the lifetimes shown in figure \ref{lifetimes} dip at certain values of $n$. The dip at $n=12$ in the S-states clearly arises from mixing with the short-lived 4d$^2$ $^3$P$_0$ state which perturbs the series at this energy. Although the corresponding channel fraction remains small (panel (b) of figure \ref{singletsstatefraction}), the partial width for the perturbing channel is much larger than that of the 5s$n$s channel (figure \ref{partialwidths}). The reasons for the dips in the lifetimes of the D states at $n=12$ and $15$ are less clear. The former occurs at the energy where the series is perturbed by the 4d$^2$ $^3$P$_2$ state; however the latter is concomitant only with the perturbation by the 4d6s $^1$D$_2$ state, which affects a broad range of D states. The results shown in panel (b) of figure \ref{partialwidths} suggest that these dips originate from how the channels interfer: the different channels of the initial D states interfere destructively in the total width, which can be seen from the fact that $\Gamma_a < \sum_i\gamma_a^{(i)}$, and the figure indicates that this destructive interference is less severe at $n=15$ and perhaps also at $n=12$, which would lead to larger decay widths for these states compared to the neighbouring ones.

The large contribution of the perturber to the natural radiative lifetimes highlights the fact that taking the perturbers into account is critical in the calculation of dipole matrix elements when considering transitions to low excited states \cite{Bergstrom1986},
owing, in particular, to the large admixture with doubly excited states in the latter.
For radiative transitions to nearby states,  however,  the contribution of these perturbers is likely to be much less significant.

\section{Conclusions}
In conclusion, we have reviewed the concepts of MQDT in its reactance matrix
formulation and developed new empirical MQDT models describing the $^1\mathrm{S}_0$, $^3\mathrm{S}_1$, $^1\mathrm{P}_1$, $^3\mathrm{P}_{0,1,2}$, $^{1,3}\mathrm{D}_2$, $^{3}\mathrm{D}_{1,3}$, $^1\mathrm{F}_3$ and $^3\mathrm{F}_{2,3,4}$ series of strontium. While these new models build upon previous work  \cite{Esherick1977,Aymar1987,Beigang1983,Armstrong1979}, they benefit from the increase in the accuracy with which the relevant energy levels have been measured in the intervening years. They are in very good agreement with experiment, and in some cases reproduce the data much more closely. We have used these MQDT models to analyse the Rydberg states considered in terms of channel fractions. Using these results, we have calculateed Land\'e $g_J$-factors for the $^{1}\mathrm{D}_2$ and $^{3}\mathrm{D}_2$ series and radiative lifetimes for the $^1\mathrm{S}_0$ and $^1\mathrm{D}_2$ series. The $g_J$-factors show good agreement with experiment, demonstrating the effect of singlet-triplet mixing in the $\mathrm{D}_2$ states of strontium. The lifetimes of the $^1\mathrm{S}_0$ and $^1\mathrm{D}_2$ states also show good agreement with the data, even for high values of the principal quantum number. The perturbers were found to lower the lifetime of the singly excited states very significantly, in particular that of the Rydberg D states, despite the smallness of their admixture in these states. 

The MQDT channel fractions and the matrix elements involving doubly excited states
found in this work  enable the calculation of Rydberg-Rydberg dipole matrix elements
in strontium taking the two-electron nature of these states into account. These dipole matrix elements are important for calculating, for example, Stark maps and dispersion coefficients \cite{Vaillant2012,Millen2011,Zhi2001}. The importance of two-electron effects in such calculations has not been ascertained to date. Their study would be a valuable addition to the current knowledge of the dynamics of Rydberg states in
two-electron atoms.

\begin{acknowledgments}
The authors wish to thank C H Greene for useful discussions. Financial support was provided by EPSRC grant EP/J007021/1 and EU grant FP7-ICT-2013-612862-HAIRS.
\end{acknowledgments}

\appendix
\section{A two-channel example}
\label{appendixA}
The simplest multichannel case is that of a two-channel model. We follow Section \ref{boundstatesection}, taking channel 1 as channel $k$ and channel 2 as channel $j$. Writing $\sigma_i = \tan(\pi\nu_i) + K_{ii}$, $i=1,2$, we obtain
\begin{equation}
\begin{vmatrix}
\sigma_1 & K_{12}\\
K_{12} & \sigma_2
\end{vmatrix}
= \sigma_1 \sigma_2 - K_{12}^2 =0.
\label{collisionexample}
\end{equation}
We can thus express $\nu_1$ as a function of $\nu_2$ in the following way: 
\begin{equation}
\nu_1 =  \frac{1}{\pi} \tan^{-1} \left( \frac{K_{12}^2}{\sigma_2} -
K_{11}\right).
\label{energyexample}
\end{equation}
The bound states energies predicted by the model are then found by determining the values of $\nu_2$ for which equation \eqref{energyexample} and the condition
\begin{equation}
I_1 -\frac{\tilde{R}}{\nu_1^2} = I_2 -\frac{\tilde{R}}{\nu_2^2}
\end{equation}
are simultaneously verified.
  
Once self-consistent values of $\nu_1$ and $\nu_2$ have been obtained,
the normalized mixing coefficients $\bar{A}_1$ and $\bar{A}_2$ can be determined by solving equation \eqref{rmatrixmqdtcondition} for $a_1$ and $a_2$, since
$\bar{A}_i = \nu_i^{3/2} a_i / \cos( \pi\nu_i)$, $i=1,2$. In this two-channel case, writing $a_2= B a_1$ yields $B = -\sigma_1/K_{12}$, and the normalization
condition (\ref{eq:normal}) requires that 
\begin{equation}
a_1 = \left[ \frac{\nu_1^3}{\cos^2 \pi \nu_1}  + \frac{\nu_2^3B^2}{\cos^2 \pi \nu_2}  \right]^{-1/2}.
\end{equation}

\section{MQDT Models}
\label{models}
The numerical values of the MQDT parameters for the models obtained in this work are presented in tables \ref{sparams} to \ref{tripletfparams}. The bound state energies predicted by these models 
can be found by proceeding as described in Section \ref{boundstatesection}, the corresponding
channel fractions as described in Section \ref{fractionssection}.
 
The uncertainties shown between brackets in the tables give an estimate of the sensitivity of the MQDT parameters to changes in the experimental energies used in the fit. They do not correspond to a full treatment of the statistical uncertainties of the parameters as these uncertainties will be highly correlated and the large number of parameters makes the interpretation of such uncertainties ambiguous. Instead, the values indicated were calculated by simultaneously shifting all the data points by one standard deviation and reoptimizing the model. The ensuing variation in the parameters gives a rough estimate of their uncertainty.

In the MQDT approach, channels are defined only by their ionization limit, 
their parity and their total angular momentum quantum number $J$. The assignments
to particular terms and configurations indicated in the tables are discussed in the 
Section \ref{SectionIII} when they might be ambiguous.

The ionization limits assumed in this work are from \cite{Esherick1977} and \cite{Beigang1982a}. For models defined in $LS$-coupling, the limits used in
 the calculations and mentioned in the tables are the
averages of the respective fine structure resolved limits.
The Rydberg constant $\tilde{R}$ was taken to be 109\, 736.627 cm$^{-1}$ \cite{Rubbmark1978}.

\begin{table*}[htbp]
\begin{ruledtabular}
\begin{tabular}{cccc}
Channel $i$ & 1 & 2 & 3\\
Ionization Limit ($\mathrm{cm}^{-1}$) & 45932.1982 & 60488.09 & 60768.48\\
\hline
Assumed channel label& $5\mathrm{s}_{1/2}n\mathrm{s}_{1/2}$ & $4\mathrm{d}_{3/2}n\mathrm{d}_{3/2}$ & $4\mathrm{d}_{5/2}n\mathrm{d}_{5/2}$\\
$K_{i 1}$ & 1.054(2) & -0.023(1) & 0.370(3)\\
$K_{i 2}$ & -0.023(1) & 2.9(3) & 0.0\\
$K_{i 3}$ & 0.370(3) & 0.0 & -0.66(1)\\
$K_{i i}^{(1)}$ & 0.83(3) & -13(10) & 0.51(7)
\end{tabular}
\end{ruledtabular}
\caption{The three-channel MQDT parameters for the $^1\mathrm{S}_0$ series of strontium. The energy dependence of the diagonals of the $K_{i \alpha}$ matrix are defined in \eqref{energydependence}. Estimates of the uncertainties in the last digit are shown in brackets.}
\label{sparams}
\end{table*}

\begin{table*}[htbp]
\begin{ruledtabular}
\begin{tabular}{ccc}
Channel $i$ & 1 & 2 \\
Ionization Limit ($\mathrm{cm}^{-1}$) & 45932.1982 &  70048.11\\
\hline
Assumed channel label& $5\mathrm{s}n\mathrm{s} \; ^3\mathrm{S}_1$ & $5\mathrm{p}n\mathrm{p}\; ^3\mathrm{P}_1$\\
$K_{i 1}$ & -34.2(4) & -155(4)\\
$K_{i 2}$ & -155(4) & -1470(40)\\
$K_{i i}^{(1)}$ & -19.14(2) & -1408(4)\\
\end{tabular}
\end{ruledtabular}
\caption{The two-channel MQDT parameters for the  $^3\mathrm{S}_1$ series of strontium. The energy dependence of the diagonals of the $K_{i \alpha}$ matrix are defined in \eqref{energydependence}. Estimates of the uncertainties in the last digit are shown in brackets.}
\label{tripletsparams}
\end{table*}

\begin{table*}[htbp]
\begin{ruledtabular}
\begin{tabular}{ccc}
Channel $i$ & 1 & 2\\
Ionization Limit ($\mathrm{cm}^{-1}$) & 45932.1982 & 60628.26\\
\hline
Assumed channel label & $5\mathrm{s}n\mathrm{p} \; ^1\mathrm{P}_1$ & $4\mathrm{d}n\mathrm{p} \; ^1\mathrm{P}_1$ \\
$K_{i 1}$ & 10.842(4)& 16.18(2)\\
$K_{i 2}$ & 16.18(2)& 22.56(3) \\
$K_{i i}^{(1)}$ & -0.39(1) & 1.68(2)\\
\end{tabular}
\end{ruledtabular}
\caption{The two-channel MQDT parameters for the $^1\mathrm{P}_1$ series of strontium. The energy dependence of the diagonals of the $K_{i \alpha}$ matrix are defined in \eqref{energydependence}. Estimates of the uncertainties in the last digit are shown in brackets.}
\label{pparams}
\end{table*}

\begin{table*}[htbp]
\begin{ruledtabular}
\begin{tabular}{cccc}
Channel $i$ & 1 & 2 & 3\\
Ionization Limit ($\mathrm{cm}^{-1}$) & 45932.1982 & 60628.26 & 60628.26\\
\hline
Assumed channel label& $5\mathrm{s}n\mathrm{p} \; ^3\mathrm{P}_0$ & $4\mathrm{d}n\mathrm{p} \; ^3\mathrm{P}_0$ & $4\mathrm{d}n\mathrm{f} \; ^3\mathrm{P}_0$\\
$K_{i 1}$ &-1.028(2) & -0.1831(3)& -0.1040(5)\\
$K_{i 2}$ &-0.1831(3) & -0.617(2)& 0\\
$K_{i 3}$ &-0.1040(5) & 0 &-0.6089(2) \\
$K_{i i}^{(1)}$ & -0.9854(5)& 0.282(4)& 0.1185(9)\\
\hline
Assumed channel label& $5\mathrm{s}n\mathrm{p} \; ^3\mathrm{P}_1$ & $4\mathrm{d}n\mathrm{p} \; ^3\mathrm{P}_1$ & $4\mathrm{d}n\mathrm{f} \; ^3\mathrm{P}_1$\\
$K_{i 1}$ &-1.063(2) &-0.1887(1) &-0.107(1) \\
$K_{i 2}$ &-0.1887(1) &-0.57(1) & 0\\
$K_{i 3}$ &-0.107(1) & 0& -0.6047(3)\\
$K_{i i}^{(1)}$ &-1.013(2) &0.09(6) & 0.083(2)\\
\hline
Assumed channel label& $5\mathrm{s}n\mathrm{p} \; ^3\mathrm{P}_2$ & $4\mathrm{d}n\mathrm{p} \; ^3\mathrm{P}_2$ & $4\mathrm{d}n\mathrm{f} \; ^3\mathrm{P}_2$\\
$K_{i 1}$ & -1.1067(9)& -0.1716(1)& -0.1273(4)\\
$K_{i 2}$ & -0.1716(1)  & -0.6052(7) & 0\\
$K_{i 3}$ & -0.1273(4) & 0 & -0.6115(5)\\
$K_{i i}^{(1)}$ & -1.0243(9)& 0.089(4)&0.086(1) \\
\end{tabular}
\end{ruledtabular}
\caption{The three-channel MQDT parameters for the $^3\mathrm{P}_0$ (top), $^3\mathrm{P}_1$ (middle) and $^3\mathrm{P}_2$ (bottom) series of strontium. The energy dependence of the diagonals of the $K_{i \alpha}$ matrix are defined in \eqref{energydependence}. Estimates of the uncertainties in the last digit are shown in brackets. The 5s13p states of all the triplet series are outliers and have been neglected, as discussed in section \ref{pstates}.}
\label{tripletpparams}
\end{table*}

\begin{table*}[htbp]
\begin{ruledtabular}
\begin{tabular}{ccccccc}
Channel $i$ & 1 & 2 &3 &4 & 5 & 6\\
Ionization Limit ($\mathrm{cm}^{-1}$) & 45932.1982 &45932.1982 & 60768.48 & 60488.09 & 70048.11 & 60628.26\\
\hline
Assumed channel label& $5\mathrm{s}_{1/2}n\mathrm{d}_{5/2} $ & $5\mathrm{s}_{1/2}n\mathrm{d}_{3/2}$& $4\mathrm{d}_{5/2} n\mathrm{s}_{1/2}$ & $4\mathrm{d}_{3/2}n\mathrm{s}_{1/2}$ & $5\mathrm{p}n\mathrm{p} \; ^1\mathrm{D}_2$ & $4\mathrm{d}n\mathrm{d} \; ^3\mathrm{P}_2$\\
$K_{i 1}$ & -0.6507(9)& -0.114(6)& -0.759(2)& $-2.7(1)\times 10^{-4}$& -0.56(1)& $-4(7)\times 10^{-4}$\\
$K_{i 2}$ & -0.114(6)& -0.489(3)& 0.362(7)& -0.644(1)& 0.11(1)& 0.0909(5)\\
$K_{i 3}$ & -0.759(2)& 0.362(7)& 1.060(2)& 0.222(1)& 0& 0\\
$K_{i 4}$ & $-2.7(1)\times 10^{-4}$& -0.644(1)& 0.222(1)& 1.172(2)& 0& 0\\
$K_{i 5}$ & -0.56(1)& 0.11(1)& 0& 0& 1.51(5)& 0\\
$K_{i 6}$ & $-4(7)\times 10^{-4}$& 0.0909(5)& 0& 0& 0&2.3660(1)\\
$K_{i i}^{(1)}$& 3.2(1)& 0.06(3)& & & 0.06(9)&\\
\end{tabular}
\end{ruledtabular}
\caption{The six-channel MQDT parameters for the $^{1,3}\mathrm{D}_2$ series of strontium. The energy dependence of the channel quantum defects are defined in \eqref{energydependence}. Estimates of the uncertainties in the last digit are shown in brackets.}
\label{evendparams}
\end{table*}

\begin{table*}[htbp]
\begin{ruledtabular}
\begin{tabular}{cccc}
Channel $i$ & 1 & 2 & 3\\
Ionization Limit ($\mathrm{cm}^{-1}$) & 45932.1982 & 60628.26 & 60628.26\\
\hline
Assumed channel label& $5\mathrm{s}n\mathrm{d} \; ^3\mathrm{D}_1$ & $4\mathrm{d}n\mathrm{s} \; ^3\mathrm{D}_1$ & $4\mathrm{d}n\mathrm{d} \; ^3\mathrm{D}_1$\\
$K_{i 1}$ & -1.55(1)& 0.549(2)& -0.0004(8)\\
$K_{i 2}$ & 0.549(2)& 1.451(1)& 0\\
$K_{i 3}$ & -0.0004(8)&0& 2.2(8)\\
$K_{i i}^{(1)}$ & -1.25(2) & 1.2(1) & -30(50)\\
\hline
Assumed channel label& $5\mathrm{s}n\mathrm{d} \; ^3\mathrm{D}_3$ & $4\mathrm{d}n\mathrm{s} \; ^3\mathrm{D}_3$ & $4\mathrm{d}n\mathrm{d} \; ^3\mathrm{D}_3$\\
$K_{i 1}$ & -1.487(8)& 0.43(2)& 0.23(2)\\
$K_{i 2}$ & 0.4(2)& 1.21(1)& 0\\
$K_{i 3}$ & 0.2(2)& 0& -0.5(2)\\
$K_{i i}^{(1)}$ & -1.10(1)& 10(2)& 7(2)\\
\end{tabular}
\end{ruledtabular}
\caption{The three-channel MQDT parameters for the $^3\mathrm{D}_1$ (top) and $^3\mathrm{D}_3$ (bottom) series of the D states of strontium. The energy dependence of the diagonals of the $K_{i \alpha}$ matrix are defined in \eqref{energydependence}. Estimates of the uncertainties in the last digit are shown in brackets. The 5s13d $^3\mathrm{D}_1$ and 5s22d $^3\mathrm{D}_3$ are outliers and have been excluded, as discussed in section \ref{odddstates}.}
\label{odddparams}
\end{table*}

\begin{table*}[htbp]
\begin{ruledtabular}
\begin{tabular}{ccc}
Channel $i$ & 1 & 2\\
Ionization Limit ($\mathrm{cm}^{-1}$) & 45932.1982 & 60628.26\\
\hline
Assumed channel label& $5\mathrm{s}n\mathrm{f} \; ^1\mathrm{F}_3$ & $4\mathrm{d}n\mathrm{p} \; ^1\mathrm{F}_3$ \\
$K_{i 1}$ & 0.383(4)&0.4522(6) \\
$K_{i 2}$ &0.4522(6) & -0.683(5) \\
$K_{i i}^{(1)}$ & 0.333(6) & -1.41(4)\\
\end{tabular}
\end{ruledtabular}
\caption{The two-channel MQDT parameters for the $^1\mathrm{F}_3$ series of strontium. The energy dependence of the channel quantum defects are defined in \eqref{energydependence}. Estimates of the uncertainties in the last digit are shown in brackets.}
\label{singletfparams}
\end{table*}

\begin{table*}[htbp]
\begin{ruledtabular}
\begin{tabular}{cccc}
Channel $i$ & 1 & 2 & 3\\
Ionization Limit ($\mathrm{cm}^{-1}$) & 45932.1982 & 60628.26 & 60628.26\\
\hline
Assumed channel label& $5\mathrm{s}n\mathrm{f} \; ^3\mathrm{F}_2$ & $4\mathrm{d}n\mathrm{p} \; ^3\mathrm{F}_2$ & $4\mathrm{d}n\mathrm{f} \; ^3\mathrm{F}_2$\\
$K_{i 1}$ & 0.6489(2)& -0.056(2)& -0.006(1)\\
$K_{i 2}$ & -0.056(2)& 1.15(6)& 0\\
$K_{i 3}$ & -0.006(2)& 0 & 2.44(2) \\
$K_{i i}^{(1)}$ & 0.4444(2)& 3.77(6)& -22.6(4)\\
\hline
Assumed channel label& $5\mathrm{s}n\mathrm{f} \; ^3\mathrm{F}_3$ & $4\mathrm{d}n\mathrm{p} \; ^3\mathrm{F}_3$ & $4\mathrm{d}n\mathrm{f} \; ^3\mathrm{F}_3$\\
$K_{i 1}$ & 0.6505(9)& -0.0512(9)& -0.007(6)\\
$K_{i 2}$ & -0.0512(9)& 1.1(2)& 0\\
$K_{i 3}$ & -0.007(6)& 0& 3.0(2)\\
$K_{i i}^{(1)}$ & 0.449(1)& 10(2)& -33(5)\\
\hline
Assumed channel label& $5\mathrm{s}n\mathrm{f} \; ^3\mathrm{F}_4$ & $4\mathrm{d}n\mathrm{p} \; ^3\mathrm{F}_4$ & $4\mathrm{d}n\mathrm{f} \; ^3\mathrm{F}_4$\\
$K_{i 1}$ & 0.662(8)& -0.03(2)& -0.011(3)\\
$K_{i 2}$ & -0.03(2)& 1.11(5)& 0\\
$K_{i 3}$ & -0.011(3)& 0 & 2.68(8)\\
$K_{i i}^{(1)}$ & 0.47(1)& 30(10)& -24.4(5)\\
\end{tabular}
\end{ruledtabular}
\caption{The three-channel MQDT parameters for the $^3\mathrm{F}_2$ (top), $^3\mathrm{F}_3$ (middle) and $^3\mathrm{F}_4$ (bottom) series of strontium. The energy dependence of the diagonals of the $K_{i i}$ matrix are defined in \eqref{energydependence}. Estimates of the uncertainties in the last digit are shown in brackets.}
\label{tripletfparams}
\end{table*}

\section{Two-electron Dipole Matrix Elements}
\label{appendixB}
In this appendix, we explain how the decay widths $\Gamma_{ba}$ defined
by equation (\ref{generalwidth}) can be written in the form of equation
(\ref{width}). We assume that the states $|\Psi_a\rangle$ and   $|\Psi_b\rangle$ 
are written in $LS$-coupling and denote by $l_{1ai}$, $l_{2ai}$, $l_{1bj}$ and $l_{2bj}$ the orbital
angular momentum quantum numbers of the two electrons in, respectively,
channel $i$ of state $a$ and channel $j$ of state $b$. Thus  
\begin{equation}
| \Psi_a\rangle = \sum_i \, \bar{A}_{ai}  |l_{1ai} l_{2ai} L_{ai} S_{ai} J_a M_{Ja} \rangle
\end{equation}
with
\begin{equation}
\begin{split}
&|l_{1ai} l_{2ai} L_{ai} S_{ai} J_a M_{Ja} \rangle = 
\sum_{M_{Sai}M_{Lai}}\sum_{m_{l{1ai}} m_{l{2ai}}}\\
 &(-1)^{L_{ai} +S_{ai}-M_{Ja} + l_{1ai} + l_{2ai} - M_{Lai}}\\ & \times \sqrt{(2L_{ai}+1)(2J_a+1)}\\
&\times \begin{pmatrix}
L_{ai} & S_{ai} & J_a\\
M_{Lai} & M_{Sai} & -M_{Ja}
\end{pmatrix}
\begin{pmatrix}
l_{1ai} & l_{ai2} & L_{ai}\\
m_{l1ai} & m_{l2ai} & -M_{Lai}
\end{pmatrix}\\
&\times |l_{1ai} m_{l1ai} l_{2ai} m_{l2ai} \rangle |S_{ai} M_{Sai}\rangle,
\end{split}
\label{decomposed}
\end{equation}
and similarly for $| \Psi_b\rangle$. The total angular momentum quantum number
$J$ and the corresponding magnetic quantum number $M_J$ are the same in
all the channels of a given state.

We take electron 1 to be inactive in the transition. Writing the dipole operator in terms 
of the spherical components of $\vec{r}_2$, namely $r_2 \sqrt{4\pi/3} \,Y_{1p}(\hat{r}_2)$, the calculation of the angular part of the dipole matrix elements reduces
to the calculation of matrix elements of the form
$$\langle l_{1bj} l_{2bj} L_{bj} S_{bj} J_b M_{Jb} | r_2 Y_{1 p}(\hat{r}_2) | 
l_{1ai} l_{2ai} L_{ai} S_{ai} J_a M_{Ja} \rangle.$$ Making use of equation
(\ref{decomposed}) and of standard sum rules \cite{Brink1993}, we find
\begin{equation}
\begin{split}
&\langle l_{1bj} l_{2bj} L_{bj} S_{bj} J_b M_{Jb} | r_2 Y_{1 p}(\hat{r}_2) | 
l_{1ai} l_{2ai} L_{ai} S_{ai} J_a M_{Ja} \rangle =\\
&(-1)^{l_{1ai}+S_{ai}  + M_{Ja}} \delta(l_{1bj},l_{1ai})\delta(S_{bj},S_{ai})
 \\ & \times \sqrt{\frac{3}{4\pi}(2l_{2ai} +1) (2l_{2bj} +1)}\\
&\times \sqrt{(2L_{ai}+1)(2L_{bj}+1)(2J_a+1)(2J_b+1)}\\
&\times \begin{pmatrix}
l_{2ai} & 1 & l_{2bj}\\
0 & 0 & 0
\end{pmatrix}
\begin{pmatrix}
J_b & 1 & J_a\\
-M_{Jb} & p & M_{Ja}
\end{pmatrix}\\
&\times \begin{Bmatrix}
J_b & 1 & J_a\\
L_{ai} & S_{ai} & L_{bj}
\end{Bmatrix}
\begin{Bmatrix}
L_{bj} & 1 & L_{ai}\\
l_{2ai} & l_{1ai} & l_{2bj}
\end{Bmatrix} R_{bj,ai},
\end{split}
\label{matrixelement2}
\end{equation}
where $R_{bj,ai}$ is the radial matrix element and $\delta$ denotes the Kronecker
$\delta$ symbol:
\begin{equation}
\delta(j,j') \equiv \delta_{jj'}.
\end{equation}
Equation (\ref{matrixelement2}) can be further simplified
by using the fact that
\begin{equation}
\begin{pmatrix}
l_{2ai} & 1 & l_{2bj}\\
0 & 0 & 0
\end{pmatrix} = (-1)^{l_{2,{\rm max}}} \left[{l_{2,{\rm max}} \over 
(2l_{2ai}+1)(2l_{2bj}+1)}\right]^{1/2},
\end{equation}
where $l_{2,\mathrm{max}} = 
\max (l_{2ai},l_{2bj})$.
Equation (\ref{width}) follows from summing the transition rate over the three components of 
$\vec{r}_2$ and over the magnetic quantum number of the final state,
which can be done using the well known formula
\begin{equation}
\sum_{p,M_{Jb}} \begin{pmatrix}
J_b & 1 & J_a\\
-M_{Jb} & p & M_{Ja}
\end{pmatrix}^2
= \frac{1}{2J_b+1}.
\label{matrixelement3}
\end{equation}

\newpage
\bibliography{mqdt}

\end{document}